%% file: main.tex
\documentclass[10pt,conference]{IEEEtran}
\IEEEoverridecommandlockouts
\input{packages_macros}

\usepackage{listings}
\usepackage{caption}
\usepackage{color}
\usepackage{datetime}
\usepackage{pifont} % in preamble

\usepackage{booktabs}
\usepackage{xcolor}
\usepackage{graphicx}
\usepackage[T1]{fontenc}

\usepackage{multirow}

\definecolor{custom-blue}{rgb}{0,0,1}

\begin{document}

\title{Propagating Developer Intent through Agentic Coordinated Refactoring}

%\title{From Developer Intent to Codebase Transformation: Multi-Agent Coordinated Rename Refactoring}
%\title{From Developer Intent to Codebase Transformation: Propagating Conceptual Changes through Coordinated Refactoring}

\author{%
\IEEEauthorblockN{Abhiram Bellur}
\IEEEauthorblockA{\textit{University of Colorado}\\
Abhiram.Bellur@colorado.edu}
\and
\IEEEauthorblockN{Mohammed Raihan Ullah}
\IEEEauthorblockA{\textit{University of Colorado}\\
raihan.ullah@colorado.edu}
\and
\IEEEauthorblockN{Dima Golubenko}
\IEEEauthorblockA{\textit{University of Colorado}\\
dima.golubenko@colorado.edu}
\and
\IEEEauthorblockN{Fraol Batole}
\IEEEauthorblockA{\textit{Tulane University}\\
fbatole@tulane.edu}
\and
\IEEEauthorblockN{Mohit Kansara}
\IEEEauthorblockA{\textit{University of Texas at Dallas}\\
mohit.kansara@utdallas.edu}
\and
\IEEEauthorblockN{Masaharu Morimoto}
\IEEEauthorblockA{\textit{NEC Corporation}\\
m-morimoto@nec.com}
\and
\IEEEauthorblockN{Kai Ishikawa}
\IEEEauthorblockA{\textit{NEC Corporation}\\
k-ishikawa@nec.com}
\and
\IEEEauthorblockN{Haifeng Chen}
\IEEEauthorblockA{\textit{NEC Laboratories America}\\
haifeng@nec-labs.com}
\and
\IEEEauthorblockN{Helena Klause}
\IEEEauthorblockA{\textit{JetBrains}\\
helena.klause@jetbrains.com}
\and
\IEEEauthorblockN{Yaroslav Zharov}
\IEEEauthorblockA{\textit{JetBrains Research}\\
yaroslav.zharov@jetbrains.com}
\and
\IEEEauthorblockN{Timofey Bryksin}
\IEEEauthorblockA{\textit{JetBrains Research}\\
timofey.bryksin@jetbrains.com}
\and
\IEEEauthorblockN{Tien N. Nguyen}
\IEEEauthorblockA{\textit{University of Texas at Dallas}\\
tien.n.nguyen@utdallas.edu}
\and
\IEEEauthorblockN{Hridesh Rajan}
\IEEEauthorblockA{\textit{Tulane University}\\
hrajan@tulane.edu}
\and
\IEEEauthorblockN{Nikolaos Tsantalis}
\IEEEauthorblockA{\textit{Concordia University}\\
nikolaos.tsantalis@concordia.ca}
\and
\IEEEauthorblockN{Danny Dig}
\IEEEauthorblockA{\textit{University of Colorado}\\
danny.dig@colorado.edu}
}

\maketitle

\input{Files/abstract}

\input{Files/introduction2}

\input{Files/motivating}
\input{Files/approach}
\input{Files/eval}

\input{Files/threats}

\input{Files/related-new}
%\input{Files/discussion}
\input{Files/conclusion}

%\section{Data Availability Statement} 
%To ensure our results are replicable and to enable future research, our experimental data, tool, prompts, and analysis scripts are publicly available~\cite{ReplicationPackage}. We will submit these artefacts for evaluation.

%\balance

\bibliographystyle{IEEEtran}

\bibliography{references,references-cia}

\end{document}

%% file: packages_macros.tex
\usepackage{graphicx}
\usepackage{textcomp}
\usepackage{xcolor}
\usepackage{amsmath,amsfonts}
\usepackage{algorithm}
\usepackage[noend]{algpseudocode}
\usepackage{array}
\usepackage[caption=false,font=normalsize,labelfont=sf,textfont=sf]{subfig}
\usepackage{textcomp}
\usepackage{stfloats}
\usepackage{url}
\usepackage{verbatim}
\usepackage{graphicx}
\usepackage{subcaption}
\usepackage{xspace}
\usepackage{hyperref}
\usepackage{listings}
\usepackage{makecell}
\usepackage{booktabs}
\usepackage{caption}
\usepackage{threeparttable}
\usepackage{longtable}
\usepackage{tabularray}
\usepackage{cleveref}
\hypersetup{hidelinks}
\usepackage{color, colortbl}
\usepackage{balance}
\usepackage[inline]{enumitem}
\usepackage{calligra}
\usepackage{amsmath,amsthm}
\usepackage{tikz}
\usetikzlibrary{shapes.geometric, arrows}
\usetikzlibrary{positioning}
\usepackage{pifont}
\usepackage{tcolorbox}
\usepackage{makecell}
\usepackage{fancybox}
\usepackage{tcolorbox}
\usepackage{graphicx}
\usepackage{caption}
\usepackage{subcaption}

\usepackage{datetime2}

\usetikzlibrary{automata, positioning, arrows.meta}

\newcommand{\dScope}{\textbf{Declared Scope}\xspace}

\newcommand{\firstAgent}{\textbf{Scope Inference Agent}\xspace}
\newcommand{\secondAgent}{\textbf{Planned Execution Agent}\xspace}
\newcommand{\thirdAgent}{\textbf{Replication Agent}\xspace}

\newcommand{\firstAgentSC}{\textsc{Scope Inference Agent}\xspace}
\newcommand{\secondAgentSC}{\textsc{Planned Execution Agent}\xspace}
\newcommand{\thirdAgentSC}{\textsc{Replication Agent}\xspace}

\newcommand{\tool}{\textsc{CoRenameAgent}\xspace}
\newcommand{\corename}{\textit{coordinated renaming}\xspace}
\newcommand{\corenameL}{\textit{Coordinated renaming}\xspace}
\newcommand{\corefactoring}{\textit{coordinated refactoring}\xspace}

\newcommand{\renas}{\textsc{RENAS}\xspace}

\newcommand{\benchSize}{{157}\xspace}
\newcommand{\totalRenameBench}{{1573}\xspace}

\newcommand{\renasBenchFilteredSize}{{161}\xspace}
\newcommand{\totalRenameRenas}{1349\xspace}

\newcommand{\avgCorenameTime}{34 minutes\xspace}
\newcommand{\avgCorenames}{5\xspace}
\newcommand{\avgFilesChanged}{4\xspace}
\newcommand{\avgLocChanged}{{150}\xspace}
\newcommand{\pctCorename}{57\%\xspace}
\newcommand{\maxCorename}{919\xspace}
\newcommand{\totalCommitsFormative}{609K\xspace}
\newcommand{\totalRenamesFormative}{249K\xspace}
\newcommand{\totalContactedDevs}{205\xspace}
\newcommand{\totalResponses}{{30}\xspace}
\newcommand{\tediousResponse}{16\xspace} % number of devs how said that corename was tedious and error prone.
\newcommand{\tediousRate}{53\%\xspace}

\newcommand{\FOneRenasBench}{{54.6}\%\xspace}
\newcommand{\PrecisionRenasBench}{{51.5}\%\xspace}
\newcommand{\RecallRenasBench}{{58\%}\xspace}

\newcommand{\FOneCORABench}{48.5\%\xspace}
\newcommand{\PrecisionCORABench}{45\%\xspace}
\newcommand{\RecallCORABench}{53\%\xspace}

\newcommand{\RenasCompletion}{{75}\xspace}

\newcommand{\bestPerformanceCORABenchFOneAlpha}{{3.1$\times$}\xspace}

\newcommand{\ablationSize}{{79}\xspace}

\newcommand{\avgCost}{{30 cents}\xspace}
\newcommand{\avgTokens}{{$\sim$77K}\xspace}
\newcommand{\avgTokensInput}{{74K}\xspace}
\newcommand{\avgTokensOutput}{{3K}\xspace}
\newcommand{\avgRuntime}{{5 minutes}\xspace}
\newcommand{\RENASruntime}{{1 hour}\xspace}

\definecolor{promptbg}{RGB}{248, 249, 250}
\definecolor{promptborder}{RGB}{209, 213, 219}
\definecolor{questioncolor}{RGB}{59, 130, 246}

\newtcolorbox{promptbox}[2][]{
    colback=promptbg,
    colframe=promptborder,
    arc=4pt,
    boxrule=1pt,
    left=10pt,
    right=10pt,
    top=8pt,
    coltitle=black,
    title={#2},
    #1
}

\tcbuselibrary{listingsutf8}
\newtcolorbox{codeboxinline}{
  boxrule=0.5pt,
  colback=gray!5,
  colframe=gray!70,
  sharp corners,
  fontupper=\ttfamily\small,
  breakable,
  enhanced
}
\newcommand{\code}[1]{{\texttt{\footnotesize{#1}}\xspace}}

\theoremstyle{definition}

\newcommand*\circled[1]{\tikz[baseline=(char.base)]{
            \node[shape=circle,draw,inner sep=1pt] (char) {#1};}}

\newcommand{\userStudyParticipantsCount}{{20}\xspace} 
\newcommand{\userStudyAvgTimesUsed}{{201}\xspace}
\newcommand{\userStudyAcceptRate}{{91\%}\xspace}

\newcommand{\userStudyAvgElapsedTime}{{7}\xspace}
\newcommand{\userStudyAvgReviewTime}{{3.5}\xspace}
\newcommand{\userStudyAvgPatternChanges}{{9}\xspace}
\newcommand{\userStudyScoreFaster}{{70\% Agree or better}\xspace}
\newcommand{\userStudyScoreReviewing}{{75\% Agree or better}\xspace}
\newcommand{\userStudyTimeSaved}{{30}\xspace}

\newcommand{\avgJavaExperience}{{1.5}\xspace}

\newcommand{\totalRenames}{{1306}\xspace}
\newcommand{\avgIdentifiersInspected}{{157}\xspace}

\newcommand{\anon}{{[Anon]}\xspace}
\newcommand{\totalProjectAnalyzed}{100\xspace}
\newcommand{\trajectoryLength}{23\xspace}

\tcbset{
  subtlebox/.style={
    colback=gray!5,
    colframe=gray!40,
    fonttitle=\bfseries,
    coltitle=black,
    boxrule=0.4pt,
    arc=3pt,
    width=\linewidth,
    halign=flush left,
    valign=center
  }
}

%% file: Files/abstract.tex
\begin{abstract}
Many software maintenance tasks require propagating a developer’s high-level intent across numerous related program elements. While recent AI agents automate software engineering tasks end-to-end, they often struggle to consistently propagate conceptual changes throughout an entire codebase. Existing approaches either rely on program analysis, which misses semantic relationships, or on large language models (LLMs), which produce incomplete or overly broad edits.

%While recent AI agents aim to automate software engineering tasks end-to-end, many practical workflows benefit from \emph{human-in-the-loop} collaboration, where developers provide high-level intent and agents perform repetitive work. This paradigm is particularly suited to tasks that require propagating developer intent for changes across related program elements. As a representative example, we study \corename refactoring, where renaming one identifier requires renaming multiple related identifiers throughout a codebase. 
%Developers perform these manually, making the process tedious and error-prone. 
%Existing heuristic approaches generate many false positives, while large language models (LLMs) produce incomplete updates.

We introduce a novel agentic workflow for \emph{propagating developer intent} through coordinated software refactoring. Our key idea is to represent a developer’s initial change as an explicit semantic specification, called a \emph{Declared Scope}, that captures the intended conceptual transformation and coordinates subsequent reasoning, execution, and propagation. 
Our workflow combines human guidance, specialized agents, and trusted IDE refactoring APIs to iteratively refine this specification and safely propagate coordinated changes across the codebase.

%We present \tool, the first multi-agent workflow for developer-guided \corename refactoring. Our key insight is that a developer’s initial rename provides a strong signal for inferring the scope of related refactorings. \tool converts this signal into a natural-language \dScope that agents use to coordinate project-wide refactorings while safely invoking trusted IDE refactoring APIs.

To understand the challenges of coordinated intent propagation, we analyzed 609K commits from 100 open-source projects and surveyed \totalResponses developers. 
Motivated by these findings, we instantiated our workflow in \tool for coordinated renaming, a representative coordinated editing task. We evaluated the approach on an established benchmark of \totalRenameRenas renames and a new uncontaminated benchmark containing \totalRenameBench recent renames. \tool substantially improves F1-score over state-of-the-art static-analysis, coding-agent, and LLM-based approaches while executing behavior-preserving refactorings through trusted IDE APIs. In a week-long user study, developers accepted \userStudyAcceptRate of the agent’s suggestions, applied \totalRenames coordinated renames, and reported saving an average of 30 minutes per refactoring session.

\end{abstract}

%% file: Files/introduction2.tex
% \todo{Compiled on \today\ at \DTMcurrenttime\ (UTC)}

\section{Introduction}
\label{sec:intro}

Software maintenance often requires developers to propagate high-level intent across many related program elements. A developer may introduce a new abstraction, rename a domain concept, migrate to a new API, or revise a design convention.
Each local edit represents a broader \emph{conceptual change} that must be propagated \emph{consistently} across the codebase.
Although such changes may appear simple, carrying them out requires developers to identify related elements, interpret project-specific conventions, and apply behavior-preserving transformations across many files. This is tedious and error-prone.
Existing IDE refactoring tools update only a selected program element and its references, but do not propagate the broader developer intent behind the change.
%Existing IDE refactoring tools provide only partial support: they reliably update a selected program element and its references, but leave developers to determine where the underlying intent should propagate next.

%This creates an opportunity for \emph{human-in-the-loop agent workflows}, where developers provide the initial intent and agents propagate it across the codebase.

%Software maintenance often requires developers to propagate concept changes (e.g., introduce a new abstraction, rename a concept, or migrate to a new API) throughout a codebase. Although these changes are seemingly simple, applying them {\em consistently} requires repeatedly locating {\em all related program elements} and updating them across multiple files. Such tasks arise in many maintenance activities, yet they remain tedious and error-prone.

%Existing tools only partially support these workflows. IDE refactoring tools operate on a single program element and its call sites at a time, leaving developers to manually identify additional affected elements. Automated approaches based on static analysis or large language models (LLMs) attempt to infer related changes but often produce large numbers of false positives or incomplete suggestions~\cite{EM-Assist,MM-Assist} (see \S\ref{sec:motivating}). This setting creates an opportunity for \emph{human-in-the-loop agent workflows}, where developers provide the initial intent and agents propagate that intent across the codebase.

% Coordinated renaming as a representative instance
This paper studies \corename as a representative coordinated editing task that requires developer intent propagation.
Coordinated renaming occurs when changing one identifier requires consistently renaming multiple semantically related identifiers to preserve conceptual consistency
For example, renaming the field \code{Customer\-Account} to \code{Client\-Account} may also require renaming the method \code{process\-Customer\-Profile} to \code{process\-Client\-Profile}.
More generally, coordinated editing tasks arise whenever a local edit expresses a broader conceptual change whose implications extend beyond syntactic references or program dependencies. Successfully automating these tasks therefore requires reasoning about developer intent and propagating it consistently across related program elements.

% Empirical motivation
To better understand how developers propagate intent across codebases, we conducted a large-scale formative study of \totalCommitsFormative commits across \totalProjectAnalyzed open-source projects and surveyed  \totalResponses developers. 
Our analysis shows coordinated renaming is common (\pctCorename of rename commits), requires substantial manual effort (\avgCorenameTime on average), and typically spans multiple refactorings across several files. Some real-world commits~\cite{intellij_commit_6d1f55f} required up to 919 coordinated renames, illustrating the scale of the problem.
%Our analysis shows that coordinated renaming is prevalent (\pctCorename of all commits with renames), time-consuming (\avgCorenameTime on average), and often requires significant manual effort to identify and update related identifiers (average of \avgCorenames rename refactorings {spread over \avgFilesChanged files}). We found examples in famous open-source projects where \corename triggered 919 distinct rename operations~\cite{intellij_commit_6d1f55f} -- indicating the scale of \corefactoring. 

These findings revealed three challenges for automation:
\begin {itemize}
\item \underline{(1) \emph{inferring and refining the intended scope}} of a conceptual change, 
\item \underline{(2) \emph{safely executing}} coordinated refactorings, and 
\item \underline{(3) \emph{propagating them consistently}} throughout the whole codebase.
\end{itemize}

%our new paradigm
To solve these challenges, we designed \tool, an agentic workflow for propagating developer intent through coordinated refactoring. A developer initiates a seed rename in the IDE, which serves as a signal of the intended conceptual change. \tool represents this intent as an explicit \dScope, iteratively refines it through lightweight developer feedback, and safely propagates coordinated refactorings across the project using trusted IDE refactoring tools. Rather than searching for, remembering, and executing every related refactoring manually, developers supervise the propagation process by reviewing suggestions and refining the inferred scope. This collaboration combines human judgment, agentic reasoning, and IDE-backed execution to substantially reduce manual effort while preserving developer control, thereby addressing a key barrier to adopting AI agents—the fear of losing control and ownership~\cite{agnia2025usingAIAssistants,kumar2025howUseAIagents}.

Existing approaches to propagating developer intent primarily follow two directions.
The first one uses program analysis, which tends to be overly strict in determining the scope of renaming. Advanced IDEs~\cite{intellij} use heuristics, e.g., renaming a class may suggest renaming test cases.
%\renas~\cite{RENAS} and \RenameExpander~\cite{renameEmpirical} 
Others~\cite{RENAS,renameEmpirical,RenameExpander} follow a change-impact-analysis workflow to expand the initial renaming and apply edits accordingly. They combine static analysis (e.g., dependency relations) with NLP-based heuristics (e.g., vocabulary similarity or structural proximity) to infer additional renames. However, as their reasoning is largely constrained by static dependencies or shallow lexical signals, they often {\bf\em fail to capture conceptual relevance among identifiers}, missing semantically related locations that should be co-renamed.
Consequently, these approaches propagate program dependencies rather than developer intent.

The second direction relies on LLMs and coding agents to infer conceptual relationships beyond static program structure. LLMs can identify semantically related identifiers within local contexts, but their reasoning degrades as prompts grow to encompass large codebases~\cite{NeedleInHaystack, shi2023large}. Recent coding agents (e.g., Codex~\cite{codex}, ClaudeCode~\cite{anthropic2025claudecode}) overcome this limitation by autonomously navigating repositories and exploring additional files. However, their exploration is largely driven by textual similarity and heuristic search, causing them to aggressively propagate renames beyond the developer’s intended conceptual change. Consequently, these agents often discover many plausible rename locations, but lack a mechanism for maintaining alignment with the developer’s intent throughout project-wide exploration.

\tool combines the strengths of both directions while addressing their limitations. Like LLM-based agents, it reasons about conceptual relationships beyond static dependencies. Like program-analysis approaches, it executes only behavior-preserving refactorings through trusted IDE tools. Unlike either approach, however, \tool continuously aligns project-wide exploration with developer intent through an explicit \dScope that is iteratively refined using developer feedback.

%To solve these challenges, {\tool} leverages a {\bf human-in-the-loop} approach that integrates LLM agents with developer feedback. It uses LLMs to propose semantically relevant rename candidates beyond strict dependency relations, while developers iteratively refine the scope through feedback, converging toward a precise yet comprehensive coordinated renaming scope.
The \firstAgent begins by analysing the \textit{seed refactoring} (i.e., a refactoring performed in the IDE or in a commit message) and transforms it into an explicit \dScope. 
%The seed refactoring originates either from a rename performed in the IDE or from a commit containing rename refactorings. 
The tool then presents renaming suggestions to the developer and collects simple go/no-go feedback, allowing \firstAgent to refine the \dScope to align with the developer's intent (\S\ref{sec:concept_agent}). 
Without developer supervision, semantic drift accumulates and propagates throughout the repository.
%Without human-in-the-loop, the agent would snowball~errors.

To address the safe execution challenge, our {\secondAgent} reasons over the \dScope within each file to identify program elements that require renaming. It presents candidates to the developer for go/no-go review and executes approved renames by invoking the IDE’s trusted refactoring tools, ensuring changes are safe by construction (see \S\ref{sec:exec_agent}). The \thirdAgent handles safe propagation through \dScope-driven file discovery, combining semantic search with program slicing to locate relevant files while avoiding unnecessary exploration (see \S\ref{sec:replication_agent}).
\tool also includes an Episodic Memory that stores developer feedback and coordinates communication among agents. \tool is implemented as an IntelliJ plugin using LangGraph~\cite{langgraph} (agentic framework) and trusted IDE refactoring APIs.

%Our framework also includes an Episodic Memory, recording developer feedback used to refine scope decisions and enables inter-agent communication. We realized our ideas in \tool, an IntelliJ IDEA plugin built with LangGraph~\cite{langgraph} (agentic framework) and powered by LLMs. After locating the corenamed elements, {\tool} invokes the IDE's refactoring engine for actual code transformations.
%OpenAI’s \texttt{o4-mini} (quantized models are often used in agentic systems for economical reasons).

%We designed an extensive evaluation of \tool to corroborate, complement, and expand research findings using multiple, complementary methods: formative study, comparative study, replication of real-world refactorings, repository mining, ablation study, user/case study, and questionnaire surveys (see \S\ref{sec:eval}).

We evaluate \tool on two benchmarks comprising 2,922 coordinated renames, including our new uncontaminated benchmark of recent commits. \tool improves F1-score by up to \bestPerformanceCORABenchFOneAlpha over the previous state-of-the-art static-analysis approach, consistently outperforms state-of-the-art coding-agent frameworks (Codex~\cite{codex} and OpenCode~\cite{opencode}), and improves up to 2$\times$ the coordinated-renaming capability of both proprietary (GPT) and open-source (Qwen) LLMs. An ablation study shows that removing developer feedback reduces precision by 4$\times$, demonstrating the importance of human supervision. Finally, in a week-long study with 20 developers, participants accepted 91\% of suggestions, applied 1306 coordinated renames, and reported saving 30 minutes per refactoring session.

Furthermore, we ensured that \tool is practical for real developers—it runs in \avgRuntime on average (compared to \RENASruntime for prior state-of-the-art) and reduces developer workload by performing \trajectoryLength actions at an affordable cost of \avgCost. While a vanilla LLM (gpt-4o-mini) can identify some related elements to rename, it produces hallucinated edits: 45–92\% of modified files fail to compile across projects, averaging 5 compilation errors per corename set. In contrast, \tool identifies most elements requiring co-renaming and executes the refactorings correctly 100\% of the time, with no compilation or semantic errors.

%CoRenameAgent --- could it be "mathematically" proven that it won't break the code itself? have you tried running test suites after applying the agent, was it all green all the time, etc.

%Finally, to validate real-world utility, 
%we used it to \todo{recommend other co-renames starting from single renames in active open-source projects. Using \tool-generated renames, we submitted \totalPRs pull requests for active open-source projects; their developers have already accepted and merged \acceptedPRs, while \rejectedPRs were rejected for socio-technical reasons and 3 are still under review}.
% \todo{Abhiram: please add a sentence on the results of the user study.}
%{we recruited \userStudyParticipantsCount participants who used it for a week on their projects. They accepted \userStudyAcceptRate of its suggestions, totalling \totalRenames applied renames.}

%Finally, to validate \tool's practical utility, we recruited \userStudyParticipantsCount participants who used it for a week on their own projects. They accepted \userStudyAcceptRate of the proposed renames, applying \totalRenames coordinated renames while saving an average of 30 minutes per refactoring session.

This paper makes the following contributions:
\begin{itemize}[leftmargin=0.7em, labelsep=0.2em]
\item \textbf{Agentic workflow for propagating developer intent} through coordinated refactoring, centered around the notion of a \dScope that explicitly represents and incrementally refines the intended conceptual change.

\item {\bf Case-study implementation.}
We instantiate this workflow in \tool, which infers the scope of co-renames and propagates them through structured reasoning and safe execution using trusted IDE refactoring APIs.

\item {\bf Empirical evaluation.}
We evaluate \tool on real-world Java projects, demonstrating substantial improvements over state-of-the-art static-analysis, coding-agent, and LLM-based approaches.

\item {\bf Benchmark.}
We release \textit{CoRenameBench}, a new benchmark of post-2025 coordinated renames that enables uncontaminated evaluation of LLM-based refactoring tools. The benchmark and all replication artifacts are publicly available~\cite{ReplicationPackage}.

\end{itemize}

%% file: Files/motivating.tex
\input{Figures/motivating_example}

\section{Motivation}
\label{sec:motivating}

\input{Files/RQs/rq1_result}

\textbf{Real World Example.}
% Now let’s dig a real-world \corename scenario.
Fig.~\ref{fig:renameagent-examples} shows a partial view of a real-world commit~\cite{flink_commit_7adeecd3} from Apache Flink. This involves coordinating large-scale renaming across hundreds of code sites, while also paying attention to subtle
semantic differences. 
% This task highlights the difficulty of changing names consistently at scale, while also paying attention to subtle semantic differences. 
% The second focuses on more selective refactoring scenarios, where subtle semantic differences require careful handling to preserve correctness in specific contexts.
% The challenge of performing large-scale renames is illustrated in a commit from Apache Flink (commit \#2839d065). 

In this commit, a developer
performed a coordinated rename from the \code{JoinHint} concept to \code{QueryHint} (\circled{1} $\rightarrow$ \circled{2}), as attested in their PR's description ~\cite{flink_pr_motivating}.
% "rename join hints to query hints". 
This coordinated change triggered renaming of 24 distinct program elements {including methods, parameters, classes, fields, and local variables. 
 The \corefactoring updated fields like \code{joinHints} into \code{queryHints},
method \code{capitalizeJoinHints} to \code{capitalizeQueryHints}, and the class 
\code{Join\-Hints\-Resolver} to \code{QueryHintsResolver}, creating 23 distinct pattern changes in names. However, the developer left public-facing methods like \code{getAllJoinHints} unchanged.
This indicates that the developer's intent was to apply the \corename \emph{only} to {internal methods} and fields.
% , leaving other identifiers untouched.
Overall, this coordinated rename affected}
% \fix{Abhiram: can we say the variety also: fields, parameters, enums, methods, etc} 
 {39} files resulting in {280} code site updates.
 
The current workflow for such large-scale, context-sensitive refactoring is tedious and error-prone. 
%A simple find-and-replace tool is too coarse, modifying every occurrence of \code{joinHints} regardless of context. 
A standard IDE ``Rename'' feature also falls short because it operates on one identifier at a time, requiring the developer to repeat the operation for all 23 related naming patterns. Manual editing is even more challenging. The developer must find the 50 files containing \code{joinHint}, inspect 585 locations, identify 24 identifiers that need renaming, and update 280 LOC. Missed renames can lead to additional code-review effort; in some cases, reviewers point out missed renames by the original author, such as PR\#25192 in Apache Flink~\cite{flink_pr_25192}.

%This process requires deep familiarity with the codebase and substantial effort to recall, search, and consistently update related identifiers. 

\textbf{Existing Approaches.}
We replicate the co-rename scenario in Fig.~\ref{fig:renameagent-examples} example with different approaches. We found that each approach struggled.
% to capture the context required for accurate renaming. 
{First, 
% we used a vanilla LLM (i.e., gpt-4o-mini). 
using a single-prompt to perform \corename is not feasible, as the vanilla LLM cannot access the whole project. 
Changes would be incomplete and confined to a single file, leading to compile errors.
% and increasing the developer's workload.
% to finish the job. 
% It suggested 1 out of the 24 renames, but it performed it incorrectly, resulting in a compile error. 
% The vanilla LLM cannot search across the project scope to discover the remaining related program elements, nor can it propagate changes across 39 files that need updates. 
% Its suggestions are confined and incomplete by nature, leaving the developer to finish the job.
}

Next, we used Claude Code~\cite{anthropic2025claudecode} with Claude Sonnet 4.6, a state-of-the-art SWE agent from Anthropic. Unlike a vanilla LLM, Claude Code can navigate project files. We prompted it to rename variables inspired by the seed rename (\code{JoinHint -> QueryHint}). Claude Code renamed 6 of the 24 elements, reaching 4 of the 39 files. We also gave Claude Code access to MCP rename refactoring tools and ran it on 10 other coordinated renames from Apache Flink. On average, it operated on 8 files and recalled 27\% of the renames with 21\% precision. These findings align with recent work~\cite{gautam2025refactorbench} showing that current SWE agents struggle with multi-file refactorings, unlike isolated function-level edits. 
%The prompts and results are available in our replication package~\cite{ReplicationPackage}.
Lastly, we used RENAS~\cite{RENAS}, a state-of-the-art research tool based on structural proximity and lexical similarity. RENAS identified only 2 of the 24 program elements to rename.

%% file: Figures/motivating_example.tex
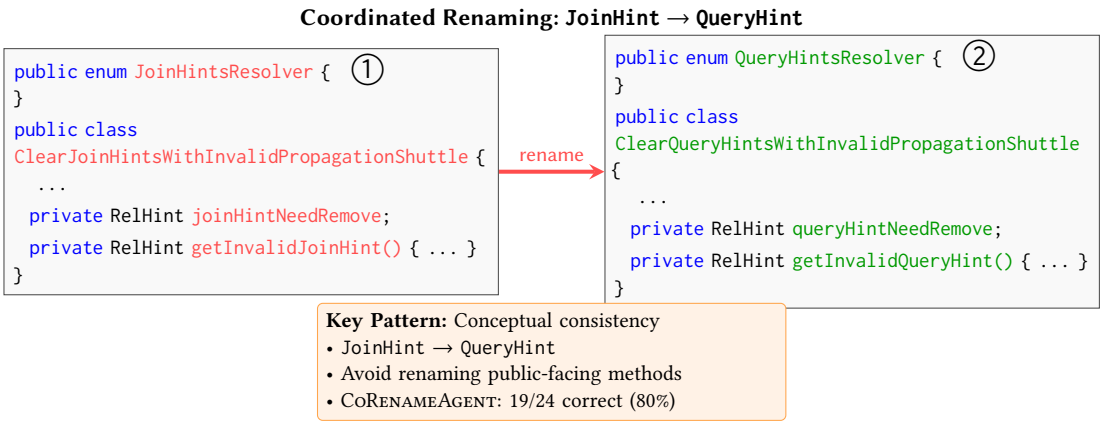
\begin{figure*}[t]
\centering
\definecolor{lightgreen}{RGB}{220,255,220}
\newcommand{\tightcircled}[1]{%
   \scalebox{1.6}{ \raisebox{0.0pt}{\textcircled{\raisebox{-0.4pt}{\scriptsize #1}}}}%
}

\begin{tikzpicture}[
    code/.style={rectangle, draw=black!70, fill=gray!5, text width=7.5cm, minimum height=1cm, font=\footnotesize\ttfamily, align=left, inner sep=4pt},
    arrow/.style={->, >=stealth, line width=1.5pt, red!70},
    label/.style={font=\small\sffamily, text=black},
    title/.style={font=\small\bfseries\sffamily, text=black}
]

\node[title] at (0,6.6) {Coordinated Renaming: \texttt{JoinHint} $\rightarrow$ \texttt{QueryHint}};

% Before (left)
\node[code] (before) at (-4.5,4.0) {
\textcolor{blue}{public enum} \textcolor{red!70}{JoinHintsResolver} \{ \tightcircled{1} \\
\} \\[2pt]
\textcolor{blue}{public class} \textcolor{red!70}{ClearJoinHintsWithInvalidPropagationShuttle} \{ \\
\hspace*{2mm} ... \\[2pt]
\hspace*{2mm}\textcolor{blue}{private} RelHint \textcolor{red!70}{joinHintNeedRemove}; \\[2pt]
\hspace*{2mm}\textcolor{blue}{private }RelHint \textcolor{red!70}{getInvalidJoinHint()} \{ ... \} \\
\}
};

% After (right)
\node[code] (after) at (4.5,4.0) {
\textcolor{blue}{public enum} \textcolor{green!60!black}{QueryHintsResolver} \{ \tightcircled{2} \\
\} \\[2pt]
\textcolor{blue}{public class} \textcolor{green!60!black}{ClearQueryHintsWithInvalidPropagationShuttle} \{ \\
\hspace*{2mm} ... \\[2pt]
\hspace*{2mm}\textcolor{blue}{private} RelHint \textcolor{green!60!black}{queryHintNeedRemove}; \\[2pt]
\hspace*{2mm}\textcolor{blue}{private} RelHint \textcolor{green!60!black}{getInvalidQueryHint()} \{ ... \} \\
\}
};

% Arrow between them
\draw[arrow] (before.east) -- node[above, font=\footnotesize\sffamily, text=red!70]{rename} (after.west);

% Key pattern box
\node[draw=orange!70, fill=orange!10, text width=6.6cm, font=\footnotesize, align=left, rounded corners=2pt] 
    at (0,6.0) {
% \textbf{Key Pattern:} Conceptual consistency\\
% • \texttt{JoinHint} → \texttt{QueryHint}\\
• Avoid renaming public-facing methods\\
• \tool: 19/24 correct (80\%)
};

\end{tikzpicture}

\vspace{-4pt}
\caption{Real-world example from Apache Flink commit 21403e31 \cite{flink_commit_7adeecd3} illustrating a \corename.}
\label{fig:renameagent-examples}
\end{figure*}

%% file: Files/RQs/rq1_result.tex
% \subsection{\bf Challenges in \corename (RQ1)}
% \subsection{\bf Formative Study}

% \paragraph{\underline{Experimental Setup}}  Our methodology involved monitoring active open-source Java projects to identify commits with significant {\corename}. 
% % {We started by identifying the most active Java projects in 2025, using the following procedure.} 
% We acquired active Java projects by using a
% % \commentR{Seart} 
% GitHub search tool~\cite{Dabic:msr2021data}, filtering for projects with more than 100 stars, 5 contributors, and 10KLOC. Then, we sorted the projects by the number of commits in 2025 and kept the top \totalProjectAnalyzed.
% % We analysed the top \totalProjectAnalyzed active Java projects. 
% We ran RefactoringMiner on all commits up to September 2025, and analysed commits that contained more {than two} \textit{coordinated renames}. 

\textbf{Formative Study.}
To understand the challenges of developers performing \corename, we analysed the top \totalProjectAnalyzed active Java projects by commit frequency. We ran RefactoringMiner~\cite{refactoringminer2} on all commits up to September 2025 and analysed commits containing more than two \textit{coordinated renames}. Our analysis of \totalCommitsFormative commits revealed \totalRenamesFormative renames, with \pctCorename co-occurring. On average, each \corename cluster contained \avgCorenames renames, changed \avgFilesChanged files, and updated \avgLocChanged LOC. Some clusters were much larger: e.g., developers performed \maxCorename renames~\cite{intellij_commit_6d1f55f} in a single intellij-community commit.

Following the firehouse survey method~\cite{hill2015spaceofbug,mazinanian2017understanding}, we monitored commit activity hourly and contacted developers who had recently performed more than five \textit{coordinated renames}. We surveyed them about time spent, perceived difficulty, and automation utility. We received \totalResponses responses from \totalContactedDevs contacted developers, who reported spending an average of \avgCorenameTime on these refactorings, ranging from a few minutes to several weeks. Most respondents (\tediousResponse/\totalResponses, \tediousRate) described the task as tedious and error-prone, citing cross-file tracking, dependency identification, and fear of breaking changes. One developer noted: \textit{``... my reliance on my memory ... meant that I left a few places where the old naming was still used.''} All of them expressed interest in automation support.

%% file: Files/approach.tex
\section{\tool}
\label{sec:approach}

\input{Figures/overview}

We designed a human-in-the-loop agentic workflow for coordinated code editing that combines LLM-based agents with human feedback to infer, refine, and propagate intended changes. Our framework consists of three specialized LLM-based agents—\textbf{scope}, \textbf{execution}, and \textbf{replication}—augmented with external tools (Table~\ref{table:agent_tools}). These agents address different stages of coordinated editing and communicate through message passing and shared episodic memory. Their prompts are dynamically updated using human feedback and outputs from other agents, allowing each stage to refine the previous one. 

{Let us follow \tool's workflow for suggesting renames on the motivating example (Fig.~\ref{fig:renameagent-examples})}. 
The workflow begins when a developer performs an initial rename operation ~(\circled{1} in Fig.~\ref{fig:overview}), called the \textit{seed rename}. 
% Instead of just replicating this edit, {\tool} deploys its first agent, 
{The {\bf \firstAgentSC} analyses this \textit{seed refactoring} in its surrounding context, to infer a broad refactoring scope --
% This agent's sole purpose is to distill the developer's underlying goal 
an explicit, evolving representation of the developer’s intended conceptual change, called the \textit{\dScope}}:
% For example, 
% For the code presented in Figure~\ref{fig:renameagent-examples}, it infers the \textit{Scope Declaration}: 
{``rename {\code{`Join'} to \code{`Query'}.}''}
This \textit{Declaration} then becomes a guide for all subsequent actions. The {\bf \secondAgentSC} uses this information to find all matching identifiers within the current file. The \secondAgentSC presents suggestions like  {\code{joinHint} $\rightarrow$ \code{queryHint}} to the developer for review ~(\circled{3}). If the developer accepts the change, \tool triggers an IDE rename operation ~(\circled{4}). If the developer rejects a suggestion
{(e.g. renaming the public method \code{getAllJoinHints})},
\tool invokes the \firstAgentSC to refine the \dScope ~(\circled{5}) --
{``avoid changes to public‐facing methods''}. 
% \commentb{For example, when the developer rejects renames of public methods such as \code{getAllJoinHints},
% the \firstAgent modifies the renaming scope -- avoiding changes to {"user‐facing methods".} }
{When the \secondAgentSC completes its execution, \tool invokes the \thirdAgentSC to propagate changes~(\circled{6}).}
To handle project-wide changes, the {\bf \thirdAgentSC} then uses the same \dScope to identify {7 other relevant files} and invokes the \secondAgentSC to perform the renames there ~(\circled{7}). 
{Finally, \tool successfully renamed 19/24 identifiers (80\%), while presenting a high rate of valid renames to the user (77\%).}
% \commentR{All three agents use an \textit{Episodic Memory} \cite{agent_arch} to store facts about their experiences {(i.e., the \dScope and the human feedback)}, and indirectly communicate with other agents. For instance, the \secondAgentSC updates the memory with review feedback from the developer, which is used by the \firstAgentSC.}
{Next, we detail the mechanisms of each agent.}

\subsection{\bf \em  \firstAgentSC (SIA)}
\label{sec:concept_agent}

%\noindent
%\begin{tcolorbox}[colback=gray!3, colframe=gray!20, boxrule=0.3pt, arc=2pt, left=6pt, right=6pt]
%\textbf{Input:} Developer feedback or the seed rename
%
%\textbf{Output:} \dScope
%\end{tcolorbox}

%**Primary function (seed → scope).**

{SIA's task is to define the refactoring scope for \tool.} 
% It is invoked (i) initially with the \textit{seed rename} to generate a \dScope\xspace{(\circled{1} in Figure \ref{fig:overview})}, and (ii) whenever the developer rejects suggestions, to 
% specialize guards
% modify the \dScope
% so future proposals avoid false positives {(\circled{5} in Figure \ref{fig:overview})}.}
This agent is our solution to the \textit{Scoping Challenge}: inferring and refining a general, project-wide rule based on  interactions with the developer. Its goal is to curate a high-level refactoring scope from the \textit{seed rename} and repeated interaction with the developer, called the \dScope. 
% The \dScope is an explicit, evolving representation of the developer’s intended conceptual change.
% \fix{define \dScope}.
{We define the \dScope $<p,g>$ as a tuple of two parts: pattern and guard.
The `pattern' $p$ specifies \emph{which} identifiers are eligible to participate in a coordinated renaming. The `guard' $g$ specifies \emph{conditions} under which the name should change. Both the pattern and guard are represented in natural language.}
\begin{itemize}
    \item $p$ is a renaming pattern of the form $\mathit{old} \mapsto \mathit{new}$, 
    \item $g$ is a guard (applicability condition) specifying the context in which $p$ should apply 
    (e.g., only to private variables).
\end{itemize}

SIA is invoked in two scenarios. (i) With the \textit{seed rename} to generate the initial \dScope\xspace{(\circled{1} in Figure \ref{fig:overview})}. (ii) When the developer rejects suggestions, to 
% specialize guards
modify the \dScope (changing the guard condition)
so future proposals avoid false positives {(\circled{5} in Figure \ref{fig:overview})}.
The prompts for \firstAgentSC are dynamic, they change each time based on the feedback from the developer.
% To curate the \dScope, the agent watches what developers do in the integrated development environment (IDE). It tracks actions like renaming (triggered in the IDE) or refactoring activity in commits, using RefactoringMiner (see table~\ref{table:agent_tools}). 
% \commentR{The agent takes the seed refactoring and the source file, then analyzes the surrounding code and produces a single, well-defined output: a natural-language \dScope. 
% {Then, as \tool asks the developer to review its suggestions, \firstAgentSC refines the scope by learning the developer's preferences. 

%We hypothesize that inferring and refining the scope addresses the primary limitation of existing tools. Using program dependencies alone does not effectively capture the ``why'' behind a \corename (which is often ambiguous and highly context-sensitive). Moreover,  separating the understanding of the scope from refactoring execution is the design principle in our approach.

% , and to manage it effectively, the agent operates in two complementary modes.

\input{Tables/list_of_tools}

{\bf Scope Extraction from Seed Rename.} 
% The \textit{seed rename} (first rename executed by the developer) indicates what identifier has been renamed, but it does not explicitly encode \textit{why} such renaming choices were made, nor reveal \emph{how} the surrounding code (semantic relationship) influenced it. Thus, to extract the \textit{Intent}, the agent treats the \textit{$r_{\text{seed}}$} as a clue. It provides an LLM with the code surrounding the edit and a structured prompt designed to produce the underlying rationale. This prompt, shown below, uses Chain-of-Thought reasoning to hypothesize about the concept. This process yields a candidate \textit{Intent} ($I$). The Concept is a natural-language rationale that explains why the seed rename was performed and how the surrounding code (e.g., function name, types) influenced that decision. We show our prompts in Figure~\ref{fig:prompt}.
The agent starts by inferring a broad and general \dScope (\circled{1} in Figure \ref{fig:overview}) by treating the \textit{seed rename} as a clue.
% , which we term as the `Generalized Scope'. 
Initially, the agent declares only the pattern for renaming (e.g., change {`join' to `query'}), leaving the guard conditions empty.
% in which the pattern is valid. 
This is a deliberate design choice to allow the scope to be refined after interactions with the developer.
We employ a few-shot example prompting strategy to extract the pattern, for its reasoning ability as it enables a contrastive analysis. 
% As shown in 
% Figure~\ref{fig:prompt}, the prompt compels the LLM to reason about where a rule should apply (Question 2), also where it should \textit{not} apply (Question 3). 
 % This fully automated mode forms the basis of our primary evaluation setup ($\alpha$ in \S\ref{subsection:rq2}), where we test {\tool}'s end-to-end performance without human intervention.

% \begin{figure}[t]
% \begin{promptbox}{Scope Refinement Prompt}
% \textbf{Given the feedback, come up with a specific condition that avoids negative reviews in the future.} \\

% Originally, I was asked to perform renames according to this pattern: 
% \\

% \{\{Pattern\}\}
% \\

% However, a few of my suggestions were rejected. See below:
% \\

% \{\{Human Review Feedback\}\}

% % \\
% % \par
% Provide a specific condition that reads like this: Do not apply this pattern in cases where ... 
% \end{promptbox}
% \vspace{-12pt}
% \caption{Scope Refinement Prompt Structure}
% \label{fig:prompt}
% \end{figure}

{\bf Scope Refinement from Developer Feedback.} 
% Complex, domain-specific logic or unique project conventions can be unclear to an LLM. Thus, to ensure the developer maintains ultimate control, we let developers to directly state or refine the \textit{Intent} $I$. 
% When a rename starts, the agent shows an optional prompt. Developers can enter their Intent, capturing details that automated analysis might miss. \commentR{We then add this human-provided rationale instead of the LLM-generated intent.} 
% By allowing direct input, we ensure that project-specific intentions are fully captured for later stages.
Complex, domain specific logic or unique project conventions can be unclear to an LLM. {To handle these cases and ensure the developer retains ultimate control, the agent refines the \dScope after interaction with the developer {(\circled{5} in Figure \ref{fig:overview})}. This developer-guided feedback enables \tool be flexible to various developers' needs. We trigger the scope refinement component when the developer rejects the suggestions. Subsequently, the \firstAgentSC modifies the guard condition in the \dScope to accommodate the developer's reviews. We achieve this by prompting the LLM with all the developers' reviews ({fetched from the episodic memory}), and ask for a condition where the rename pattern should \textit{not} be applied. This is done to ensure that LLM does not hallucinate and restrict the scope too harshly, preventing suggestions in the future.
Finally, the new \dScope is stored back in memory for the other agents to access. 
Prompts are in the replication package~\cite{ReplicationPackage}.
% create controlled experimental setups ($\beta$ and $\gamma$ in \S\ref{subsection:rq2}) where we can analyze the downstream impact of a perfect or human-assisted scope, thereby isolating the performance of the execution and propagation agents.
}

{\bf Walkthrough: \firstAgentSC.} In the example from Figure~\ref{fig:renameagent-examples}, SIA reviews the \textit{seed rename} and yields a \dScope: `Perform renames following this pattern: {join -> query}'. 
After the developer rejects the suggestion to rename the {public method ``getAllJoinHints'' to ``getAllQueryHints''}, the \firstAgentSC refines the \dScope by adding the guard condition - ``Do not apply this rename when the declaration you are touching is a public method for backward compatibility reasons''.

% {\em When a method signature has a generic parameter named `function' but serves a specific role, especially alongside a counterpart method like `left'—the developer should rename it to a domain-specific, role-reflecting name (e.g. `rightFunction') to mirror symmetry, clarify intent, and prevent ambiguity in the API}.
%\vspace{-7pt}

\subsection{\bf \em \secondAgentSC (PEA)}
\label{sec:exec_agent}

%\noindent
%\begin{tcolorbox}[colback=gray!3, colframe=gray!20, boxrule=0.3pt, arc=2pt, left=6pt, right=6pt]
%\textbf{Input:} Source code, \dScope \\
%\textbf{Output:} Rename suggestions (for developer to review)
%\end{tcolorbox}

The goal of the \secondAgentSC (PEA) is to analyse a file to suggest and execute renames that align with the \dScope. 
% The {\secondAgentSC}'s candidate identification uses a ReAct-style loop $(think \rightarrow act \rightarrow observe \rightarrow plan)$, bootstrapped with few-shot exemplars drawn from memory. 
% The task of the \secondAgent is to take the \dScope and source file as input, and suggest renames to the developer -- finally executing approved refactorings. 
% In order to do so, the \secondAgent must translate the high-level \dScope \commentb{(pattern + guard conditions)} into a sequence of concrete, safe refactoring operations. 
The PEA addresses two challenges:
% This translation presents two fundamental challenges: 
(1) accurately identifying which program elements match the \dScope, and (2) ensuring that every {modification preserves the program's behaviour.} 
% We address these by designing an agent that combines reasoning with tool-based actions.

% {\bf Candidate Identification.}
% \label{sec:candidate_indentification}
% The first challenge is determining which identifiers in a file match the \dScope. We design the agent around the ReAct framework~\cite{yao2023react}, which enables the \secondAgentSC to reason like a developer: decide what to check, invoke a tool, observe the result, and plan the next step. We further guide the agent with task-specific few-shot examples constructed from accepted/rejected suggestions stored in the \textit{Episodic Memory}. These examples reflect the developer's current scope and help the agent avoid repeated mistakes. We use a reasoning model to identify renaming candidates, as such models are effective when guided by clear objectives~\cite{yao2023react}. In this step, the \secondAgentSC produces renaming suggestions, including the identifier name, type, new name.

{\bf Candidate Identification.}
\label{sec:candidate_indentification}
The first challenge is determining which identifiers in a file match the \dScope. We design the agent around the ReAct framework~\cite{yao2023react}, enabling the \secondAgentSC to reason like a developer: decide what to check, invoke a tool, observe the result, and plan the next step. We further guide the agent with task-specific few-shot examples from accepted/rejected suggestions in the \textit{Episodic Memory}. These examples reflect the developer's scope and avoid repeated mistakes. We use a reasoning model to identify renaming candidates, as such models are effective when guided by clear objectives~\cite{yao2023react}. In this step, the \secondAgentSC produces 
renaming suggestions, including name, type, and new name.

%The first challenge is determining which identifiers within a file match the \dScope. To address this, we designed the agent around the ReAct framework~\cite{yao2023react}, with additional guidance provided by few-shot examples generated on the fly. ReAct allows the \secondAgentSC to mimic the methodical process of a developer: thinking about what to check, using a tool to execute it, observing the result, and then planning the next step. We enhance the agent's ability to reason by providing it with task-specific few-shot examples. We construct these examples based on the history of accepted/rejected suggestions stored in the \textit{Episodic Memory}. These examples are specific to the current refactoring task and exemplify the developer's scope. This ensures that the agent reasons effectively. We use a state-of-the-art model with reasoning capabilities for identify renaming candidates, as recent work suggests such models excel at reasoning when guided by a clear objective~\cite{yao2023react}.  In this step, the \secondAgentSC  produces a series of renaming suggestions (i.e, list of identifier name, identifier type, new name, and line number).

% The agent's task is to produce a plan consisting of a list of `(element\_name, element\_type, new\_name)' tuples, effectively deciding both \textit{what} to change and \textit{which tool} to invoke. 

{\bf Refactoring Execution.}
The second challenge is ensuring execution safety. A naive approach would be to let an LLM directly edit the code. However, a core risk with LLMs is their tendency to hallucinate, producing code edits that are buggy
% or syntactically plausible but semantically incorrect
~\cite{EM-Assist, MM-Assist, dilhara2024unprecedented}. This exposes the developer to unpredictable and potentially dangerous modifications.
Therefore, to mitigate this risk, our design enforces a critical safety constraint -- the agent is strictly prohibited from directly editing the source code. Instead, starting with the suggested rename candidates, \secondAgentSC 
executes changes by invoking a curated set of IDE-backed refactoring tools:
% e.g. \texttt{rename\_variable(variable\_name, new\_name)}.
% performs a series of validation steps described below:

% As the agent is provided with multiple refactoring execution tools (see Table \ref{table:agent_tools}), it has the autonomy to trigger the appropriate tool(s). We supply two tools to the agent:

% \begin{enumerate}
    (1) Rename identifier: rename any type of identifier - method, class, local variable, field. This tool runs safety checks before executing changes. If the agent attempts an invalid rename, a descriptive error message is returned to the agent.
    
    (2) Update comment: a find \& replace tool which updates the existing comments in the source code. This is crucial because the IDE cannot propagate renames in comments. Thus, the agent invokes a find \& replace tool on the comments only, as an LLM is capable of understanding the natural language content of a comment.
% \end{enumerate}

{\bf Validation and Adjustment.}
To execute changes, the agent must convert rename suggestions into valid refactoring tool calls~(\ref{sec:candidate_indentification}). However, LLMs may hallucinate the element location, such as an incorrect line number, or the element type, such as labeling a field as a variable. 
To address this, we build a lightweight validation wrapper that makes a best-effort attempt to locate the target identifier.

Given a list of (\textit{identifier\_name}, \textit{identifier\_type}, \textit{new\_name}, \textit{line\_number})) from the LLM, the wrapper produces valid IDE rename calls by adjusting the line number and identifier type when needed. It inspects the file, finds declared and referenced identifiers matching the suggested name, and selects the target based on type and closeness to the expected line. If no element is found, it reports an error; if multiple elements match, it chooses the closest one by line number.

{\bf Safe
% \commentR{Verifiable} 
Execution.} The rename tools provided to the agent wrap the IDE's internal APIs, leveraging its AST and type-checking system. Thus, refactorings are safe by construction: invalid operations, such as renaming a nonexistent method, fail with descriptive errors returned to the agent for re-planning. The IDE also checks semantic preconditions before applying changes, such as preventing variable shadowing or preserving method dispatch in inheritance hierarchies. To allow renames in any order, \tool identifies each target by line number; because renaming does not change line positions, these references remain stable. \tool applies a rename only when all preconditions are satisfied.
%The rename tools provided to the agent are wrappers around the IDE's internal APIs that leverage its Abstract Syntax Tree (AST) and type-checking system. Therefore, this tool-mediated approach guarantees that every refactoring is safe by construction, as any invalid operation (e.g., renaming a method that does not exist) will simply fail with a descriptive error that is returned to the agent for re-planning. {Crucially, these tools perform sophisticated static analyses to ensure semantic correctness before applying any transformation. Even for an apparently simple operation like renaming, the IDE enforces sophisticated preconditions to guarantee safety. For instance, it checks that a new name does not introduce variable shadowing, and that renaming a method does not inadvertently override or disrupt method dispatch in the inheritance hierarchy. To ensure renames can be applied in any order, we identify each identifier by its line number. Because renaming does not change the number or positions of lines, these references remain stable, allowing renames to be applied in any sequence. \tool proceeds with a rename operation only when all preconditions are satisfied.

{\bf Human Feedback.}
The \secondAgentSC presents validated and safe suggestions to the developer inside the IDE, allowing them to accept or reject each suggestion (see~\circled{3} in Fig.~\ref{fig:overview}). We assume the developer understands the intended conceptual realignment, either because they initiated it or discussed it during design or code review, and can judge whether a suggestion is valid. The \secondAgentSC stores the developer's feedback and code snippets in memory for other agents. Based on rejections, the \firstAgentSC updates the scope (\circled{5} in Fig.~\ref{fig:overview}), and \tool safely executes accepted suggestions in the IDE (\circled{4} in Fig.~\ref{fig:overview}).

%The \secondAgentSC presents validated and safe suggestions to the developer inside their IDE, offering an option to accept or reject them {(see ~\circled{3} in Fig. \ref{fig:overview})}. We assume that the human is aware of the required conceptual realignment and understands its details, either because they initiated it or discussed it with peers during design or code review. Thus, the developer is a reliably judge, and can determine whether a given renaming suggestion is valid. After receiving human feedback, the \secondAgentSC then stores the developer's feedback (accept/reject decisions and code snippets) in memory for use by other agents. Based on rejections, the \firstAgentSC analyses the feedback and adjusts the scope {(\circled{5} in Figure \ref{fig:overview})}. Finally, the \tool executes accepted suggestions safely in the IDE {(\circled{4} in Fig. \ref{fig:overview})}.

{\bf Termination.}
Finally, the \secondAgent's loop terminates ~(\circled{6} in Figure~\ref{fig:overview}) when one of the following is satisfied: (i) the agent \textit{reasons} that there are no more necessary renames, (ii) the iteration limit has been reached, 
(iii) there are no new suggestions, i.e., all the suggestions in the current iteration have already been offered before,
(iv) or there are three consecutive failing tool calls.

% \fix{Abhiram: Attempted fix, please review. this is one example of a clear place where we need to beef it up for OOPSLA reviewers. They would care a lot more about semantic, correctness, etc. We can expand here that even seemingly simple refactorings like renames, have sophisticated program analyses to determine that when we rename we don't accidentally overshadow variables, that we don't accidentally override methods in in inheritance class hierarchy, etc. That's why it's crucial to invoke a high-semantics tool like a refactoring tool which has a much deeper understanding of programming language constraints than an LLM. Such checks are encoded in the refactoring preconditions as static analysis checks.}

{\bf Walkthrough: \secondAgentSC.} In the motivating example in Figure~\ref{fig:renameagent-examples}, the execution agent, starts by examining {\code{JoinHintsResolver.java}}. The agent reasons that the local field {\code{allOptionsInJoinHints}} needs to be renamed to {\code{allOptionsInQueryHints}}, and the method \code{initJoinHints} needs to be renamed to \code{initQueryHints}. It proceeds to obtain approval from the developer, and then executes these rename using the IDE tools. Then, it reasons that "there are no other program elements that need to change according to the provided scope". The execution agent then stops its execution.

%\vspace{-10pt}
\subsection{\bf \em \thirdAgentSC (RA)}
\label{sec:replication_agent}

%\noindent
%\begin{tcolorbox}[colback=gray!3, colframe=gray!20, boxrule=0.3pt, arc=2pt, left=6pt, right=6pt]
%\textbf{Input:} Executed Renames \(\tilde{C}_{\mathcal{T}}\) \\
%\textbf{Output:} List of source files to inspect
%\end{tcolorbox}

{The \thirdAgentSC (RA) has one  task: 
% given a set of changes, 
given previously edited sites {(\circled{6} in Figure \ref{fig:overview})},
discover new files to edit according to the \dScope}. 
{RA follows two steps: 1) Discovery (edited sites $\rightarrow$ new files), and 2) Filtering. Finally, RA passes discovered files to the {\secondAgentSC} for further analysis {(\circled{7} in Figure \ref{fig:overview})}.
% \thirdAgentSC combines structural reachability from edited sites (slicing/call-graph navigation) and semantic search (keywords) to discover new files; then it filters each candidate to yield the next set of files to inspect.
}

% {The \thirdAgentSC is tasked with propagating the inferred \dScope across the entire project, {given previously edited sites {(\circled{6} in Figure \ref{fig:overview})}}. Its goal is to produce a candidate set of files where the refactoring may be applicable, which are then passed to the {\secondAgentSC} for further analysis {(\circled{7} in Figure \ref{fig:overview})}. 
The central challenge for the \thirdAgentSC is to discover a focused set of candidate files where renames should propagate. 
A brute-force analysis of every file is computationally infeasible and would overwhelm downstream agents. 
Instead, the task decomposes into identifying files that are: (1) structurally coupled to the initial changes, and (2) conceptually related but structurally decoupled.
{To address these challenges, the \thirdAgentSC employs a two-pronged strategy:
% that combines static analysis with a heuristic-based technique. 
(i) it uses program slicing to analyze structural dependencies, and (ii) it performs a targeted keyword search to discover semantically-related files. We detail these steps below.}

% {This task is challenging because relevant files can be linked either structurally (e.g., via direct method calls) or purely semantically (e.g., by sharing naming conventions in unrelated modules). A naive approach, like analyzing every file, is computationally infeasible and would overwhelm downstream agents. }

% \commentb{We rely on this insight to find relevant files - renames are often a consequence of high-level semantic changes (e.g. type changes). As a result, identifiers that are changed together are often related via data flow (forwards or backwards).}

% \paragraph{\bf \commentR{Step 1: Structural Traversal via Call-Graph Analysis}} 
% \commentD{Comments on formalism}
% \commentb{First, the agent begins with the set of renames that were performed by the \textit{Planned Execution Agent}. Using the IDE's static analysis infrastructure (see Table~\ref{tab:agent_tools}), we traverse the call-graph forwards and backwards -- following the data flow of renamed identifiers. In our implementation of \tool, we invoke IntelliJ's forward and backwards `Data Flow Analysis'. This process yields an set of files that are structurally coupled to the renaming change.}

{\bf {Step 1.1: Discovering Structurally-coupled Files via Program Slicing.}}
We rely on the insight that renames often follow high-level semantic changes, such as type changes. Thus, identifiers changed together are often connected through data flow. To identify structurally coupled files, we use inter-procedural program slicing.
The process starts from the set of renames already executed by the tool, including the original seed rename performed by the developer and the additional elements modified by the \textit{Planned Execution Agent} in the initial file. We use these renames as a multi-point slicing criterion to discover statements coupled through data or control dependencies. For each program element, we compute both its backward slice, i.e., statements that affect it, and its forward slice, i.e., statements affected by it. The candidate files for \dScope replication are those containing at least one structurally related statement. For scalability, slicing is bounded by recursion depth and timeout.

%We rely on the insight that renames are often a consequence of high-level semantic changes (e.g., type changes). As a result, identifiers that are changed together are often related via data flow. Therefore, to identify structurally-coupled files, we use inter-procedural program slicing. {The process begins with the set of renaming operations previously executed by the tool. Initially, this set comprises of the original seed rename performed by the developer and all subsequent program elements modified by the \textit{Planned Execution Agent} in the initial file. We use the executed renames as a multi-point slicing criterion to discover all statements coupled through data or control dependencies.

%For each program element, we compute the backward slice (statements that affect it), and the forward slice (statements that are affected by it) to produce a set of structurally related statements. The candidate files for replication of the \dScope are the files containing at least one structurally related statement. For scalability, program slicing is bounded by recursion depth and a timeout.

% This method provides a structural basis to identify relevant files, ensuring that any code directly coupled by data or control dependencies is included for analysis.

{\bf {Step 1.2: Discovering Semantic Candidates via Keyword Search.}} 
{Program slicing may not capture all relevant files, especially those linked by high-level concepts or coding conventions rather than direct data flow. Our insight is that files within the same package are often changed together. To this end, \thirdAgentSC searches for keywords (based on the history of accepted renames) within the previously modified packages and aligned test directories.
For example, if the developer accepts {\code{capitalizeJoinHints} to \code{capitalizeQueryHints} in the file \code{src/main/com/tool/File.java}, the agent uses the keyword \code{capitalizeJoinHints}} to search within the directories \code{src/main/com/tool}, and \code{src/test/com/tool}.}

{\bf {Step 2: Filtering.}} {After the above steps, the agent is left with a set of unique candidate files. To avoid the cost and noise of running the full execution process on every file, it performs a crucial filtering step. For each candidate file, the agent asks the following question: \textit{``Given the \dScope and the candidate file, should the refactoring be replicated in \textit{this} file?''}.\xspace If the answer is yes, \tool invokes the \secondAgentSC to analyse that file.}

{The \secondAgentSC and \thirdAgentSC work hand-in-hand until a fix point is reached, iteratively discovering more files to inspect, and applying rename refactorings in those files. The \secondAgentSC applies renames, and the \thirdAgentSC inspects the codebase for other relevant files. The loop is terminated when a fix-point is reached -- i.e, the \thirdAgentSC finds no new files to inspect. For practical reasons, we terminate this fix-point algorithm after three iterations. }

{\bf Walkthrough: \thirdAgentSC (RA).} On the motivating example Figure~\ref{fig:renameagent-examples}, the RA iteratively retrieves {61} related files. After using the LLM for filtering, we are left with {26} relevant files. The \textit{PEA} then performs {38} refactorings.

\subsection{\bf \em Episodic Memory}
\label{sec:memory}

% What is episodic memory~\cite{agent_arch}. 
An episodic memory~\cite{agent_arch} is a short-lived memory that persists only during one execution of \tool and is refreshed at the start of each run. It stores facts from the current workflow, including developer interactions, past actions, and information needed for inter-agent communication.

\tool uses this memory to store developer feedback on suggested renames, i.e., accept or reject, and the current \dScope. The \firstAgentSC uses the feedback to refine and update the \dScope. The \secondAgentSC (PEA) constructs few-shot examples from past feedback to guide its reasoning and avoid repeated suggestions. It also uses memory to keep the LLM context focused and prevent prompts from growing unnecessarily across iterations. Finally, the \thirdAgentSC uses the stored \dScope to discover related files.

%An episodic memory~\cite{agent_arch} is a short-lived memory that persists only during a single execution of \tool, and is refreshed at the start of each new run. This memory stores facts about the agent's experiences during the current workflow. We use this kind of memory for three reasons: (i) To remember the history of interaction with the developer, (ii) remember actions taken in the past, and (iii) to facilitate inter-agent communication. The memory enables on-the-fly adaptation of \tool to the developer’s preferences.  \tool stores the following information in memory: the developer's feedback for suggested renames (accept/reject), and the current \dScope. The \firstAgentSC uses the stored developer feedback to refine and update the \dScope in memory. The \secondAgentSC (PEA) constructs few-shot examples from past feedback to guide its reasoning.  These examples serve as a reminder to PEA, preventing repeated suggestions. PEA uses the memory to minimally construct the LLM’s context -- focusing attention on relevant text and preventing the prompt from growing unnecessarily with each iteration. Finally, the \thirdAgentSC leverages the \dScope stored in memory to discover related files.

{\bf Implementation. }{\tool's implementation details can be found in our companion technical report \anon{}}.

%% file: Figures/overview.tex
\begin{figure*}[htpb]
    \centering    
    \includegraphics[width=0.9\linewidth,trim=0cm 0cm 0cm 0cm]{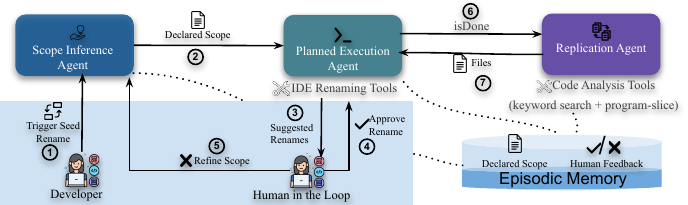}
     \vspace{-3pt}
    \caption{Workflow of \tool}
    \label{fig:overview}
\end{figure*}

%% file: Tables/list_of_tools.tex
%\vspace{-5pt}
\begin{table}[t]
\small
\footnotesize
  \centering
  \caption{Agent‐Level Tools for Co‐Rename Pipeline.}
  \label{table:agent_tools}
  \vspace{-3pt}
  \tabcolsep 3pt
\begin{tabular}{|c|l|}
\hline
\textbf{Agent}      & \textbf{Tool and Description}     
% & 
% \textbf{Description}                                   
\\
\hline
\makecell{Scope Inference }  & Monitor IDE Renames: Detect rename in the IDE    
% &  Detect cases when developers trigger renames in the IDE 
\\
 (\ref{sec:concept_agent}) & Monitor commit actions -- Pull latest commits
 % & Detect cases when developers manually refactored the code  
 \\
 
 & RefactoringMiner -- Detect refactorings in a commit
 % & Detects manually applied refactorings in a commit   
 \\ \hline
                    % & Precondition Checker  & Verifies semantic constraints (type correctness, visibility, override relations). \\
\makecell{Planned Execution } 
& Rename: Parameter, Local Variable, Method, Field, Class
% & Rename a method's parameter, including signature, call sites.
\\
% (\ref{sec:exec_agent}) & Rename Local Variable 
% % & Rename local variables and their references within its scope.      
% \\
%                     & Rename Method         
%                     % & Update method, references, overridden/overriding relations.  
%                     \\
%                     & Rename Field         
%                     % & Change field name in class definition and all references.     
%                     \\
%                     & Rename Class        
                    % & Change class name in its declaration, update references.    
                    
                    (\ref{sec:exec_agent}) & Update Comment         
                    % & Update an existing comment in the file to reflect the renaming.   
                    \\
                    \hline
\makecell{Replication (\ref{sec:replication_agent})}& {Program Slicing} 
-- Find related locations of an element
% {Perform a forwards and backwards program slice of a given program element, find relevant locations in the project.}
\\ 
                    & {Keyword Search} --
                    Finds files containing keywords 
                    % related to the \dScope, beyond direct dependencies.      
                    \\ \hline

\end{tabular}
\end{table}

%\vspace{-19pt}

%% file: Files/eval.tex
\section{Empirical Evaluation}
\label{sec:eval}

% We set out to evaluate {\tool} by answering the following research questions:

We designed a comprehensive, multi-methodology evaluation of \tool to corroborate, complement, and expand
research findings: benchmark-based evaluation (with replication of real-world refactorings), ablation analysis, operational-cost measurement, and a user study. Our evaluation addresses four questions:

\begin{enumerate}[label=\bfseries RQ\arabic*.,wide, labelwidth=!, labelindent=0pt]

\item \textbf{Effectiveness.} How accurately does \tool identify
coordinated renames compared with LLM-only baselines,
general-purpose coding agents, and state-of-the-art
program-analysis tools?

\item \textbf{Ablation.} Which architectural components are
necessary for propagating developer intent across a project?

\item \textbf{Operational Cost.} What are the monetary,
latency, token, and interaction costs of running \tool?

\item \textbf{Usefulness.} Do developers trust \tool,
and does supervising the workflow reduce manual effort?

\end{enumerate}

\input{Tables/renas_dataset}

\subsection{\bf Subject Systems}

% \todo{Abhiram add details about the subject systems}

We evaluate \tool\ on two Java datasets to assess both reproducibility (RenasBench [5]) and generalizability (our new Co-renameBench). We use the RENAS benchmark to enable direct comparison with prior work. However, because its commits predate the training cut-offs of modern LLMs, they may have been memorized during pretraining~\cite{dong2024generalization}, 
% introducing data contamination and 
threatening the validity of LLM-based evaluation.
% (data contamination).

To mitigate this threat, we created Co-renameBench, a new benchmark mined from GitHub commits after May 2025, well beyond the May 2024 training cut-off of OpenAI's \texttt{o4-mini}. We used RefactoringMiner~\cite{RefactoringMiner} to detect rename refactorings, then two authors manually grouped developer-performed renames into \corename sets.

Finally, we applied three filters to focus on non-trivial coordinated renames. First, we removed overriding-method renames, which existing IDEs already automate. Second, we retained only renames spanning multiple files. Third, we retained only instances containing at least 4 renames, 
%Our formative study showed that commits with fewer than 4 renames were typically independent renames, whereas larger clusters more often reflected coordinated conceptual changes.
based on findings from our formative study where commits with fewer than 4 renames are typically independent renames. 
%, while ≥4 renames often reflect a co-rename  pattern. 
%because smaller cases are typically manageable using existing IDE support. 
% Our analysis of 609K commits shows long-tail distribution of renames/commit (median = 2, mean = 5), with most commits involving only a few renames. 
% We find that commits with fewer than 4 renames are typically independent renames, while 4 or more renames often reflect a co-rename  pattern. 
The resulting benchmarks contain 161 RenasBench instances and 157 uncontaminated Co-renameBench instances (see Table~\ref{table:renas_dataset}). We ran experiments on a 16-core MacBook with 16 GB RAM.

\input{Files/RQs/rq2_result}
\input{Files/RQs/rq3_result-new}
\input{Files/RQs/rq4_result}
\input{Files/RQs/rq5_result}

%% file: Tables/renas_dataset.tex
\begin{table}[H]
% \small
\footnotesize
\centering
\caption{Evaluation datasets used in our study}
\vspace{-6pt}
\tabcolsep 2.6pt
\label{table:renas_dataset}
\begin{tabular}{llrrrr}
\toprule
\textbf{Benchmark} & \textbf{Name}    & \textbf{\# files} & \textbf{LOC} & \textbf{\# Renamings} & \textbf{\# Sets} \\
\midrule
\multirow{3}{*}{RenasBench~\cite{RENAS}} & RatPack         & 831                    & 42K       & 916                 & 
% 327              
{109}
\\
                           & ArgoUML         & 1884                  & 175K      & 443                   & 
                           % 178              
                           {52}
                           \\
                           & \textbf{Subtotal}  & 2715                  & 217K      & \totalRenameRenas                 & \renasBenchFilteredSize   %505
                           \\
\midrule
\multirow{3}{*}{Co-renameBench} & Flink  & 13K           & 1.6M  & 340          & 39      \\
                                & Kafka  & 5.5K            & 920K  & 211          & 13       \\
                                & Spring-Boot  & 7.7K            & 474K  & 83          & 12       \\
                                & IntelliJ-Community  & 82K            & 4.9M  & 85          & 10       \\
                               & Camunda  & 12K            & 1.0M  & 582           & 60       \\
                               & Bytechef  & 4K            & 244K  & 145          & 15       \\
                              & MekHQ  & 1.4K            & 286K  & 127          & 8       \\

                               & \textbf{Subtotal}  & 125.6K            & 9.4M  & \totalRenameBench           & \benchSize       \\
\midrule
\textbf{} & \textbf{Total}  & 128K            & 9.6M   & 2771           & {318}       \\
\bottomrule
\end{tabular}

\textit{Note: K = thousands, M = millions}
\end{table}

%% file: Files/RQs/rq2_result.tex
\subsection{\bf Effectiveness of \tool (RQ1)}
\label{subsection:rq2}

%\subsubsection{{\color{custom-blue}RQ1.1. Improving an underlying LLM}} We used X foundation LLMs: GPT-4o, GPT-5 (?), and Qwen-2.5-Coder-14B (?).

%1. GPT-4o-mini:

%\begin{itemize}
%    \item GPT-4o-mini: F1\_score =   22\%
    
%     \item GPT-4o-mini + {\tool} : F1\_score = 51.5\%

%\end{itemize}

%2. GPT-5:

%\begin{itemize}

%\item GPT-5: F1\_score = ??

%\item GPT-5 + {\tool} : F1\_score = ??

%\end{itemize}

%3. Qwen-2.5-Coder-14B:

%\begin{itemize}

%\item Qwen-2.5-Coder-14B: F1\_score = ??

%\item Qwen-2.5-Coder-14B + {\tool}: F1\_score = ??

%\end{itemize}

%\subsubsection{{\color{custom-blue} RQ1.2. Agentic Framework Comparison}} We configured Codex~\cite{codex} with the underlying model GPT-5 (?) and compared its corenaming effectiveness with our {\tool} framework using the same GPT-5 model.

%GPT-5 + {\tool} > GPT-5 + Codex

%\subsubsection{{\color{custom-blue} RQ1.3. Comparison with State-of-the-Art Corenaming Tools}}

%Comparison with RENAS

%\begin{itemize}
%    \item GPT-4o-mini + {\tool} > RENAS
%    \item Qwen + {\tool} > RENAS
%\end{itemize}

\subsubsection{Baseline Tools}
We compare \tool\ against three classes of approaches. First, we evaluate whether our workflow improves the coordinated-renaming capability of underlying foundation models, both proprietary (o4-mini) and open-source (Qwen3.6 35B-A3B), chosen for their favorable cost-performance tradeoff. Second, we compare against state-of-the-art coding agents, OpenCode~\cite{opencode} and Codex~\cite{codex}. We configured OpenCode with the Eclipse JDT Language Server~\cite{jdtls} and the same underlying models (o4-mini and Qwen3.6) to isolate the effect of the workflow rather than the LLM. We evaluated Codex with GPT-5.4-mini; its o4-mini configuration consistently failed to modify the codebase, so we omit those results. Third, we compare against RENAS~\cite{RENAS}, the previous state-of-the-art static-analysis approach, which is representative of traditional change impact analysis tools based on program-dependencies. For reference, we also report results from vanilla prompting using GPT-4o-mini, GPT-5-mini, GPT-5, o4-mini, and Qwen3.6.

\subsubsection{Evaluation Metrics}
We consider a predicted rename to be a true positive (TP) if it correctly identifies a program element from the gold-set as a target for renaming.
A rename is considered as a false positive (FP) if an unrelated identifier is erroneously flagged, and is not in the gold-set. 
All renames that are part of the gold-set but were missed by a tool are considered as false negatives (FN). 
We measure effectiveness using standard Precision, Recall, and F1-score with respect to the benchmark's gold-set. 
% Precision (TP / (TP + FP)) quantifies the proportion of correct renames among all suggestions, while Recall (TP / (TP + FN)) captures the fraction of actual renames uncovered. F1-score is the harmonic mean of both. 
Note that Precision measures whether the correct identifiers are selected for coordinated renaming, not whether the underlying rename transformation preserves behavior. Transformation correctness is guaranteed by the IDE refactoring engine. %, which safely executes every approved rename.
We compute all metrics on the entire \corename set (i.e., over the entire dataset).
% If there are many cases where either precision or recall is high while the other is low, this leads to an average F1-score that is lower than either the average precision or average recall. For a practical tool, it is important that both precision and recall are high and balanced, thus
% we will use F1-score as the key performance indicator.

 %Harmonic mean is dominated by the smaller value, so if Precision and Recall are imbalanced, F1 suffers. Averaging F1 scores (e.g., in macro-F1 or per-instance F1) highlights worst-performing segments, whereas Precision and Recall averages may hide them by dilution. 

%Thus, we report `Average Precision', `Average Recall', `Average F1-Score' in our tables

% \input{Figures/eval_setup}

\subsubsection{Experimental Setup} 
We applied the same evaluation protocol to both RenasBench and Co-renameBench. Each experiment begins from a developer-initiated seed rename and evaluates how accurately each approach propagates that intent throughout the project.

Following \renas~\cite{RENAS}, each coordinated rename set is initiated from a single seed rename chosen using the priority Class > Field > Method > Parameter > Local Variable, reflecting the rename developers are most likely to perform first. We exclude the seed rename from evaluation.

%{Our evaluation setup {borrows from} the \renas authors:
%for each co-rename set (gold-set) in the ground truth, we execute one of the renames based on a priority order {(Rename Class > Field >~Method > Parameter > Local Variable)} to serve as the \textit{seed rename} for \tool. 
%{The priority order reflects the rename operation likely to be performed first by the developer. }
% {Next, we execute the seed rename, and pass it to \tool as input to perform \corename.
%While evaluating approaches, we remove the \textit{seed rename} from the gold-set.}

Developer supervision is evaluated only for \tool because it is an integral part of its algorithm. To provide a fair comparison, we additionally evaluate state-of-the-art coding agents that, like \tool, perform iterative multi-step execution, thereby separating the benefits of autonomous agentic execution from those of developer supervision.

Because \tool\ relies on developer supervision, we model the developer using a binary oracle that accepts or rejects each proposed rename. The oracle never reveals additional gold-set entries or future renames, providing only the same local feedback available to a human reviewer. Consequently, repository exploration, reasoning, and intent propagation remain entirely the responsibility of the agent—the developer merely evaluates discovered suggestions rather than helping discover new ones. The oracle validates suggestions but never guides the search. This setup enables large-scale, reproducible evaluation while faithfully modeling developer supervision.

{To ensure fair
% possible 
comparison with \renas, beyond running the publicly available tool with default settings, we directly corresponded with RENAS's authors to validate that our~setup 
produced the correct output. Further, we validated our setup of \renas by reproducing the precision and recall presented in their paper.
% numbers the \renas authors presented in their paper, validating our experimental setup.
%{As the \renas tool also produces results for \RenameExpander, we report those as generated.} 
% \commentb{\renas fails to execute on some inputs, primarily due to unsupported Java syntax in later versions. To ensure a fair comparison, we exclude these failed cases when computing performance metrics. For instance, in Table~\ref{table:combined_results}, \renas successfully completed only \RenasCompletion instances; therefore, we report its precision and recall only for those \RenasCompletion cases, without assigning zero values for the \RenasIncomplete instances where execution failed.}
We run \tool with LLM~temperature of 0 to maximize determinism. %We used OpenAI's o4-mini.

\input{Tables/rq2_table}

{

\subsubsection{Results} 
Table~\ref{table:combined_results} shows that \tool consistently achieves the highest F1-score across both benchmarks. 
The gains hold across three complementary classes of baselines—static analysis, coding agents, and vanilla LLMs—indicating that explicit intent propagation, rather than any individual reasoning capability, is the missing ingredient.
The three baseline categories reveal complementary limitations: static analysis lacks semantic reasoning, coding agents lack explicit intent representation, and vanilla LLMs lack project-scale coordination. \tool addresses all. % three limitations.
% through \emph{explicit intent propagation}.

%Moreover, the results show that our agentic workflow improves the co-rename capability of the underlying LLM.

%Table~\ref{table:combined_results} presents the effectiveness of \tool across both datasets. Overall, the results show that our framework improves the co-rename capability of the underlying LLM, outperforms general-purpose coding-agent frameworks in most settings, and achieves higher F1-score than the state-of-the-art program-analysis-based approach.

These results suggest that the central challenge is not generating rename candidates, but maintaining alignment with developer intent during repository-wide propagation. In large projects, semantically related identifiers often span packages, layers, or modules that are difficult for LLMs to connect reliably. Moreover, the intended propagation depends on project-specific conventions—for example, whether public APIs should be renamed—which cannot be inferred from code alone. By iteratively refining the \dScope through lightweight developer feedback, \tool adapts to evolving constraints while propagating changes.

%\commentb{Our findings show that the key challenge in coordinated renaming is repo-level navigation. In real-world, large, and diverse projects, co-renamed elements often span multiple packages or application layers. These semantic links, as well as project-specific preferences, are difficult for an LLM to capture reliably. Coordinated renaming also depends on developer intent: for example, rapidly evolving projects may allow public API renames, whereas stable projects may avoid them. Such preferences are often ambiguous without developer feedback. }

%\commentb{\tool adapts to each refactoring session by learning from developer feedback and refining the declared scope.
%Across both benchmarks, \tool may still miss valid renames when the \firstAgentSC over-constrains the scope, especially when feedback contains too few positive examples. Inferring guard conditions from limited or imbalanced feedback is challenging~\cite{APIFix,pyevolve} and remains future work. In practice, it mitigates this issue by exposing guard conditions for developer review and modification.}

\paragraph{Boosting the co-rename capability of an LLM}
Vanilla LLMs already identify semantically related identifiers within a file, leading to high precision, but they achieve low recall because they cannot navigate and reason across files in the full project. By traversing the repository, \tool\ improves recall by more than 2$\times$ for both o4-mini and Qwen. More capable LLMs may reason more accurately about individual rename candidates, but coordinated refactoring additionally requires maintaining a persistent representation of developer intent throughout repository exploration. Our workflow supplies this capability independently of the underlying model.
Future advances in LLM reasoning are complementary rather than substitutive: stronger models improve individual decisions, while explicit intent propagation coordinates those decisions consistently across the repository.
The architectural benefit is therefore orthogonal to advances in LLMs.

% As shown in Table~\ref{table:combined_results}, \tool substantially improves the performance of its base model. With o4-mini, \tool improves F1-score from 28\% to 54.6\% on RenasBench and from 25\% to 48.5\% on Co-renameBench. Similar improvements were observed when using {\tool} with Qwen3.6. We also compare \tool with vanilla LLMs. The best model-only F1-score is 29\% on RenasBench and 33\% on Co-renameBench, while \tool with o4-mini achieves 54.6\% and 48.5\%, respectively.
{%Vanilla LLMs already identify semantically related identifiers within a file, but they struggle to propagate the same conceptual change across an entire repository.}

\paragraph{Comparing with frontier coding-agents--Codex and OpenCode}
General-purpose coding agents successfully navigate repositories, but their precision deteriorates as exploration expands beyond the initial file because they lack an explicit representation of the developer’s intended conceptual change. Although OpenCode and Codex perform repository-wide exploration, they increasingly exhibit \emph{semantic drift}, proposing semantically plausible identifiers that fall outside the intended rename scope. In one representative example, OpenCode (with o4-mini) applied a project-wide scripted replacement changing \code{doc} to \code{document}, producing over 3,700 unnecessary renames despite achieving higher recall.  

This semantic drift becomes even more pronounced as exploration moves beyond the seed file. OpenCode starts with 60\% precision on the initial file but drops to 7\% (i.e., 8.5$\times$ lower) on secondary files when using o4-mini. Likewise, Codex (with GPT-5.4-mini) falls from 53\% to 24.5\% (2$\times$ lower). In contrast, \tool\ maintains substantially more stable precision across files (55\% → 40\%) by explicitly representing the \dScope and refining it through developer feedback rather than allowing the intended conceptual change to drift during exploration.  

These results suggest that repository navigation and iterative execution alone are insufficient for coordinated refactoring. General-purpose coding agents can effectively explore repositories, but they lack a task-specific mechanism for inferring, refining, and consistently propagating developer intent. Explicit intent propagation—not simply stronger agents—is therefore the key to accurate coordinated refactoring.

%\commentb{Observations:
%- Agents do a lot of keyword searching, which will lead to things outside of the "concept".

%- Agents tends to "create the concept" out of thin air -- i.e., just by looking at the code. They do not look backwards in the commit history, or use other signals. This leads to "generic" catch-all concepts, giving them a high recall, at the cost of precision.

%\todo{This is comparable to our setup in Table IV, row 1, replication On.}
%}

% Compared with OpenCode, \tool with o4-mini achieves higher F1-score using the same base model on both RenasBench (54\% vs. 41\%) and Co-renameBench (48.5\% vs. {16}\%). Similarly, \tool with Qwen outperforms OpenCode with Qwen on Co-renameBench (41\% vs. 37\%), although it is slightly below OpenCode on RenasBench (44\% vs. 46\%).

%\commentb{Compared with general-purpose coding agents, \tool produces results that better align with the developer’s intended rename scope (Table~\ref{table:combined_results}). Codex and OpenCode rely heavily on keyword-based exploration, which often pulls in identifiers that are only loosely related to the target concept. In one case, OpenCode (with o4-mini) applied a project-wide scripted replacement (via shell and regex-based edits) to change \texttt{doc} to \texttt{document}, resulting in 3700 unnecessary renames and a sharp drop in precision, despite of having higher recall. In contrast, \tool constrains the search using an explicitly defined and refined scope, leading to more precise and consistent renaming.
}

\paragraph{Comparing with static-analysis approaches}
Compared with RENAS~\cite{RENAS}, the state-of-the-art static-analysis approach, \tool\ improves F1-score by 2.3$\times$–3.1$\times$. Static analysis excels at propagating renames along explicit program dependencies such as references, call graphs, or inheritance relations, but coordinated renaming is fundamentally a \emph{conceptual} rather than \emph{structural} task. Semantically related identifiers often span packages, architectural layers, or naming conventions without sharing direct control-flow, data-flow, or call-graph relationships.

For example, renaming \code{CustomerAccount} to \code{ClientAccount} may also require renaming \code{processCustomerProfile}, even though the two identifiers might not be directly connected in the program graph. Whether such propagation is appropriate depends on the developer’s intended conceptual change. By explicitly representing and refining this intent through the \dScope, \tool\ propagates coordinated renames beyond syntactic relationships while remaining under developer supervision.

\vspace{-5pt}
\begin{tcolorbox}[
  colback=gray!5,
  colframe=gray!60!black,
  boxrule=0.8pt,
  arc=3mm,
  left=4mm,
  right=4mm,
  top=0.5mm,
  bottom=0.5mm,
  fonttitle=\bfseries,
]

% \tool cons
Neither stronger LLMs, coding agents, nor static analysis alone solve coordinated refactoring. Explicit propagation of developer intent is the missing capability.
\end{tcolorbox}
%\vspace{-5pt}

% \subsubsection{\todo{\bf Overlap analysis}}

% \paragraph{\underline{Experimental Setup}} We evaluate overlap on the full RENAS benchmark of rename tasks (RatPack and ArgoUML; see Table~\ref{table:renas_dataset}). For each task, we collect the co-rename sets produced by \tool and the baseline tools. We then count, per task, how many suggestions each tool shares with each other tool and how many are unique to each. Aggregated totals are visualized in \todo{Figure~\ref{fig:venn_diagram}} as a Venn diagram.

% \paragraph{\underline{Results}} As illustrated in \todo{Figure~\ref{fig:venn_diagram}}, the tools demonstrated interesting complementary patterns. \tool and RENAS agreed on \todo{X\%} of their suggestions, primarily for cases where identifiers were directly connected within the code. However, each tool also identified unique renaming opportunities. Specifically, RENAS found \todo{Y} renames that \tool overlooked, while \tool discovered \todo{Z} renames that RENAS missed. For instance, \tool identified \todo{XYZ}, which is a connection that structural analysis alone can miss. In contrast, RENAS found structurally linked identifiers that \tool sometimes overlooked. {\color{orange}{Tien: We need to characterize the strengths of our tool over RENAS.}} These findings indicate that our scope-driven approach effectively complements existing structural methods. \todo{What if we combine both approaches?}

%\vspace{-0.15in}

%% file: Tables/rq2_table.tex
\begin{table}[H]
\centering
\footnotesize
\caption{ Evaluation results of \tool compared to baselines. RENAS* ran without errors on \RenasCompletion CorenameBench entries; metrics are computed only on those runs (no penalty for failures on others).
} 
\label{table:combined_results}
\tabcolsep 2.5pt
\vspace{-6pt}
\begin{tabular}{l|ccc|ccc}
\toprule
& \multicolumn{3}{c|}{\textbf{RenasBench}} & \multicolumn{3}{c}{\textbf{Co-renameBench}} \\
\cmidrule{2-7}
\textbf{{Tool}} & \textbf{Prec.} & \textbf{Rec.} & \textbf{F1} & \textbf{Prec.} & \textbf{Rec.} & \textbf{F1} \\
\midrule
% \textbf{static analysis} \\
\renas* [static analysis]   & 14\% & {73\%} & {24}\% & 10\% & 35.3\% & 15.6\% \\

\midrule
% \textbf{LLMs} \\

GPT-4o-mini [LLMs] & 31\% & 17\% & 22\% & \textbf{71\%} & 10\% & 18\% \\

GPT-5-mini & 50\% & 19\% & 27\% & {60\%} & 16\% & 25\% \\

GPT-5.4-mini & {56\%} & {19\%} & {28\%} & {51\%} & {16\%} & {24\%} \\

{GPT-5} & \textbf{53\%} & 17\% & 26\% & 60\% & 22\% & 33\%\\

{o4-mini} & {52\%} & 19\% & 28\% & 46\% & 17\% & 25\%\\

{Qwen3.6} & {59\%} & {19\%} & {29\%} & 59\% & 17\% & 26\%\\

\midrule
% \textbf{Agents}\\
{Codex w/ gpt-5.4-mini} [Agents] & {36\%} & \textbf{83\%} & 50\% & {28}\% & {\textbf{73}}\% & {41}\%\\

{OpenCode w/ o4-mini} & {38\%} & {45}\% & {41}\% & {10}\% & {43}\% & {16}\%\\

{OpenCode w/ Qwen3.6} & {35\%} & 67\% & 46\% & {25}\% & {71}\% & 37\%\\

\midrule

\tool w/ Qwen  & {44\%} & {44\%} & {44\%} & 42\% & 40\% & 41\% \\

\tool w/ o4-mini  & {\PrecisionRenasBench} & \RecallRenasBench & \textbf{\FOneRenasBench} & \PrecisionCORABench & {\RecallCORABench} & \textbf{\FOneCORABench} \\

\bottomrule
\end{tabular}
\end{table}

% \begin{tablenotes}
% \small
% \centering
% % \item \todo{This table merges Table 3 and Table 4. I thought a single table instead of two would be easier to read, and we also save space. What does everyone think? If it looks better, I will remove table 4 and 5.}
% \end{tablenotes}
%\vspace{-5pt}

%% file: Files/RQs/rq3_result-new.tex
\subsection{\bf Ablation Study (RQ2)}

\begin{table}[t]
\centering
\footnotesize
\setlength{\tabcolsep}{2.5pt}
\caption{Ablation study on \textit{Co-renameBench}. We incrementally add the replication agent, human feedback, and scope refinement to the same base model, o4-mini.}
\label{ablation_table}
% \begin{tabular}{lcccc}
% \toprule
% \textbf{Configuration} & \textbf{Agentic} & \textbf{Human} & \textbf{Scope} & \textbf{F1} \\
% & \textbf{Loop} & \textbf{Feedback} & \textbf{Refinement} & \\
% \midrule
% Vanilla LLM (o4-mini) & No & No & No & 25\% \\
% o4-mini + agentic loop & Yes & No & No & 19\% \\
% o4-mini + agentic loop + human & Yes & Yes & No & 30\% \\
% Full \tool{} & Yes & Yes & Yes & 51.5\% \\
% \bottomrule
% \end{tabular}
\begin{tabular}{lccc}
\toprule
\textbf{Configuration} & \textbf{Precision} & \textbf{Recall} & \textbf{F1} \\
\midrule
o4-mini & 46\% & 17\% & 25\% \\
o4-mini + replication & 11\% & 64\% & 19\% \\
o4-mini + replication + human feedback & 20\% & 63\% & 30\% \\
Full \tool{} & 47\% & 56\% & 51.5\% \\
\bottomrule
\end{tabular}
\end{table}

{

\subsubsection{Experimental Setup}
We designed the ablation study to evaluate whether each architectural component addresses a distinct failure mode in propagating developer intent. Starting from a vanilla LLM (o4-mini), which reasons only within the file of the seed rename, we incrementally enable the Replication Agent, human feedback, and Scope Refinement. We evaluate each configuration on a subset of Co-renameBench containing \ablationSize coordinated renaming instances.

\subsubsection{Results}
Table~\ref{ablation_table} reveals a clear progression. Project-wide exploration dramatically increases recall by discovering coordinated renames beyond the initial file, but also introduces semantic drift that causes precision to collapse. Human feedback and Scope Refinement progressively restore precision by keeping propagation aligned with developer intent, ultimately yielding the highest F1-score.

The Replication Agent overcomes the vanilla LLM’s inability to propagate developer intent beyond its local context. By exploring semantically related files, it propagates the inferred \dScope throughout the project, increasing recall from 17\% to 64\%. However, this aggressive exploration introduces \emph{semantic drift}, proposing plausible but out-of-scope renames and reducing precision by 4$\times$. 
% \todo{say that precision drop as we go further and further away from the starting file.} 
This shows that naive repository exploration is not sufficient for \corename.

Human feedback corrects this semantic drift without limiting project-wide exploration. Precision doubles by preventing semantically misaligned, out-of-scope renames from propagating further, while recall remains high. However, feedback alone corrects individual suggestions without updating the inferred scope, allowing similar mistakes to recur.
% elsewhere in the project.

Finally, Scope Refinement generalizes developer feedback by refining the \dScope with guard conditions that guide subsequent exploration, substantially improving precision. However, Scope Inference Analysis (SIA) occasionally over-constrains the scope, filtering out valid rename candidates and slightly reducing recall.

\begin{tcolorbox}[
colback=gray!5,
colframe=gray!60!black,
boxrule=0.8pt,
arc=3mm,
left=4mm,
right=4mm,
top=0.5mm,
bottom=0.5mm,
fonttitle=\bfseries,
]
The ablation validates our architectural decomposition: Replication enables project-wide exploration, while human supervision and Scope Refinement are essential to prevent semantic drift and keep propagation aligned with developer intent.
%Developer supervision is essential to detect errors before they propagate. Without human feedback, precision degrades by 4X.
\end{tcolorbox}

}

%% file: Files/RQs/rq4_result.tex
\subsection{\bf Operational Cost Analysis (RQ3)}

{\underline{Experimental Setup.}} We quantify \tool’s operational overhead per refactoring session using LangSmith to track: 
(1) \emph{Monetary cost} in USD, (2) \emph{Token consumption} (input and output), and (3) \emph{Latency} (end-to-end time). 
We monitored \tool (setup with \texttt{OpenAI o4-mini}) while conducting experiments over our benchmarks.

{\underline{Results.}} 
We find that across all refactoring tasks in our benchmarks, the average operational cost was \avgCost per task while consuming approximately \avgTokens tokens. The mean end-to-end latency was \avgRuntime, consisting of Scope Extraction (10 seconds), file discovery (1.5 minutes), and planning and execution (3 minutes). Despite performing structured project-wide reasoning, \tool\ consumes 10–20$\times$ fewer tokens than general-purpose coding agents (Table~\ref{tab:cross_setup_comparison}). This efficiency stems from its task-specific workflow, which narrows exploration using the inferred \dScope instead of repeatedly reasoning over a the repository.

\begin{table}[t]
\centering
\begin{tabular}{lcccc}
\hline
\textbf{Approach} & \textbf{Tokens (In -- Out)} & \textbf{Steps} & \textbf{Time} \\
\hline
Codex (GPT-5.4-mini) & 1.48M -- 14.9K & 14 & 2 min \\
OpenCode (o4-mini)   & 336.7K -- 1.9K & 19 & 1.5 min \\
OpenCode (Qwen3-6)   & 783.8K -- 5.7K & 17 & 2 min \\
RENAS      & -- & -- & {60 min} \\

\tool (o4-mini)      & \textit{\avgTokensInput -- \avgTokensOutput} & \textit{\trajectoryLength} & {5 min} \\
\hline
\end{tabular}
\caption{Operational costs of various approaches.}
\label{tab:cross_setup_comparison}
\end{table}

{
% The state-of-the-art static analysis tool \renas requires an average of \RENASruntime to analyze the same refactoring tasks. 
{Latency differences with RENAS can be attributed to
% The difference stems from their 
fundamentally different approaches: \renas exhaustively computes program dependencies and lexical similarities across the entire codebase upfront, making it suitable only for batch processing during off-hours.} 
% Moreover, on large projects with 1M+ LOC this approach scales poorly, demanding a huge heap memory.
% , otherwise it will terminate with out of memory exceptions.
In contrast, \tool's {scope-driven}, \emph{on-demand} approach explores only the relevant portions of the codebase. 
With \tool's \avgRuntime latency, developers can delegate the heavy lifting to it and spend minimal time reviewing high-quality suggestions. 

% \commentR{\tool presents \avgSuggestions candidate renames for review on average. 
% This hybrid approach strikes a balance: it offloads the tedious tasks of searching and propagating renames while preserving human judgment for oversight. 
% By amplifying developers' effectiveness rather than replacing them, \tool enables a more natural, cooperative refactoring workflow.}

% With \tool's \avgRuntime response time, developers can delegate their work to \tool, and spend their time reviewing it's suggestions.
% iterate on conceptual renames as naturally as they would with simple find-and-replace operations, 
% but with the semantic awareness that ensures correctness across the project. 
% {Thus, enabling interactive refactoring within the developer's workflow.}}
% \fix{Discuss about human alone, tool alone, vs hybrid approch's qualitative cost.}

% \todo{Danny says: let's now report the running time for RENAS, then we can say we perform better and we execute in much less time. Thus our solution can be used in an interactive scenario, whereas previous solutions require a batch processing mode.} -- Thanks for the feedback. It's revised based on your suggestion.
\vspace{-5pt}
\begin{tcolorbox}[
  colback=gray!5,
  colframe=gray!60!black,
  boxrule=0.8pt,
  arc=3mm,
  left=4mm,
  right=4mm,
  top=0.5mm,
  bottom=0.5mm,
  fonttitle=\bfseries,
]
%\tool is fit for practical development -- on average, it costs \avgCost per invocation.
Explicit intent propagation not only improves accuracy but also makes repository exploration 10–20$\times$ more token-efficient than frontier coding agents.
% workflows: it is affordable, fast, IDE-integrated, and does not overwhelm developers with too many false positives.
\end{tcolorbox}
%\vspace{-10pt}

%% file: Files/RQs/rq5_result.tex
\subsection{\bf Usefulness Study (RQ4)}

%Beyond correctness on benchmarks, a crucial measure of a tool's value is 
%To assess our tool's utility, we used two methods: a user study with students, and patches to open-source~projects.
% an ``in-the-wild'' study submitting pull requests to open-source projects.

%\subsubsection{Experimental Setup}

{\underline{Experimental Setup.}}
To assess \tool’s practical utility, we recruited \userStudyParticipantsCount graduate students (average \avgJavaExperience years of Java experience) who used the tool for one week on their own Java projects. Participants shared anonymous telemetry capturing fine-grained usage statistics, including accepted suggestions and review time. At the end of the study, they completed a 6-point Likert-scale survey measuring perceived suggestion quality, supervision effort, and estimated time savings. The survey instrument and responses are available in our replication package~\cite{ReplicationPackage}.

{\underline{Results.}}
Over the one-week study, participants invoked \tool\ \userStudyAvgTimesUsed times. Participants consistently trusted its project-wide propagation, accepting \userStudyAcceptRate of the proposed coordinated renames, resulting in \totalRenames applied renames. Each session generated 8 suggestions on average, keeping supervision lightweight. Participants refined the inferred \dScope only \userStudyAvgPatternChanges times across all sessions, indicating that the inferred scope correctly captured the intended change.

Developer supervision required relatively little effort. On average, \tool executed for \userStudyAvgElapsedTime minutes per session, while developers spent only \userStudyAvgReviewTime minutes reviewing suggestions. During that time, the agent analyzed 11 files and inspected \avgIdentifiersInspected unique identifiers to produce just 8 coordinated rename suggestions spanning 5 files, hiding the complexity of repository-wide exploration behind a lightweight supervision interface. Participants estimated that this saved an average of \userStudyTimeSaved minutes compared to performing coordinated renaming manually. Moreover, \tool was faster than manual refactoring (\userStudyScoreFaster), and  the required level of supervision was appropriate (\userStudyScoreReviewing).

%We also analyzed the manual effort required to use the tool. On average, the agent required \userStudyAvgElapsedTime minutes to complete execution, while participants spent \userStudyAvgReviewTime minutes reviewing its suggestions. Participants tolerated the runtime: the agent ran to completion in {185 out of \userStudyAvgTimesUsed} invocations. For this effort, the agent analysed 11 files and reviewed \avgIdentifiersInspected unique identifiers to suggest 8 renames across \userStudyAvgFiles files. 
%Survey responses indicate that using the agent was faster than performing renames manually (\userStudyScoreFaster), and that it saved an average of \userStudyTimeSaved minutes. Participants also felt they performed an appropriate amount of reviewing (\userStudyScoreReviewing). 
% When asked to estimate the approximate time saved per refactoring task, participants reported an average savings of \userStudyTimeSaved minutes.

In survey responses, participants consistently highlighted \tool’s ability to \emph{propagate conceptual changes across a project}. One participant reported that it ``saved significant time by finding related terms'', while another described using it to transform singular identifiers into plural ones throughout a project—a coordinated edit they found difficult to perform manually. Compared with frontier coding agents, participants noted that \tool\ prioritized accurate, developer-supervised suggestions over speed.

%{In survey responses, participants highlighted \tool's practical strengths. One mentioned ``saved significant time by finding related terms''. Another said they transformed singular to plural names, a surprising capability. Reflecting on frontier coding agents, participants said \tool's suggestions arrived slower.}
% compared to other coding agents. 

%\subsubsection{In-the-Wild Study}
%We monitored active projects for incomplete \corename (1-2 renames in a commit) to serve as seeds for \tool. Then, we submitted \totalPRs pull requests.

%{\underline {Accepted Contributions.}} Maintainers merged \acceptedPRs of our PRs. 
%For example, in \textsc{Kafka}, our tool propagated a variable rename across three classes (5 renames total), which the maintainer merged after confirming it followed project conventions.

%{\underline {Learning from Rejections.}} Rejections (\rejectedPRs) highlighted socio-technical factors. 
%In one case, developers declined a PR because the review effort was thought to outweigh the change's benefit.
%In another, the rename conflicted with subtle project-specific stylistic conventions, underscoring the need for the developer in the loop.

\vspace{-5pt}
\begin{tcolorbox}[
  colback=gray!5,
  colframe=gray!60!black,
  boxrule=0.8pt,
  arc=3mm,
  left=0.5mm,
  right=0.5mm,
  top=0.5mm,
  bottom=0.5mm,
  fonttitle=\bfseries,
]
Developers trusted \tool’s project-wide propagation, accepting \userStudyAcceptRate of suggestions while spending only 3.5 minutes supervising each session, saving an average of 30 minutes per coordinated refactoring.
\end{tcolorbox}

% \tool to reason not just about code but also about project-specific styles and the perceived value of a change.

%% file: Files/threats.tex
\section{Threats to Validity}
\label{sec:threats}

%\subsubsection*
%{\bf Construct Validity.} Our study uses precision, recall, and F1-score as proxies for the effectiveness of \corename. While these metrics do not capture all aspects of a refactoring's quality, they are standard in refactoring literature~\cite{RENAS,dong2024context,zhang2023accurate}. We use the same evaluation metrics as the state-of-the-art baselines, ensuring that our comparisons are direct and fair. Similarly, we use API token counts and latency to measure \textbf{cost}, which are established proxies for computational and financial overhead in LLM-based systems.

%\subsubsection*
\textbf{Internal Validity.} To mitigate the non-determinism of LLMs, we executed each refactoring using a fixed model and temperature of~0.

\textbf{External Validity.} We evaluated \tool exclusively on open-source Java projects, which may limit the generalizability of our findings to other programming languages or proprietary codebases. 
However, the core multi-agent framework of \tool (i.e., separating Scope Extraction from tool-driven propagation) is language-agnostic. 
To adapt \tool to another language like Python, one would need to replace the Java-specific components (rename and program-slicing implementations) while the fundamental agentic workflow remains the same. 
% \fix{Talk about linked PRs/commits causing oracle incorrectness.}

{\bf Dataset Validity.}
% \commentb{Our dataset validity hinges on the fact that developers perform \corename thoroughly. This may not always be the case. For example, developer may initial miss to perform some rename, but do it in a later commit/PR, when pointed out by a reviewer. Future work may expand such datasets by finding other commits with related reanaming activity, thereby expanding the \corename sets.}
{We build our benchmarks assuming that developers perform \corename comprehensively in a single commit. In practice, some renames may be delayed or distributed across multiple commits, for example, when a reviewer later points out an omission. 
{As a result, our oracle derived from commits may be incomplete, leading to false negatives (i.e., valid but unobserved renames), though false positives are unlikely given the use of mature, reviewed repositories. Consequently, suggestions not present in the oracle may be penalized despite being semantically valid.}
Future work could
link related commits to capture a complete view of \corename activity, thereby expanding the oracle.}

%% file: Files/related-new.tex
\section{Related Work}
\label{sec:related}

We group related work into: (1) Change Impact Analysis, (2)
suggesting Rename Refactoring, 
% discovering \corename, 
and (3) LLM-based Refactoring Agents.
% , and (4) coding agents.

{\bf Change Impact Analysis}. 
Change Impact Analysis (CIA) identifies parts of a system affected by a code change, supporting testing and maintenance~\cite{DBLP:journals/tosem/RothermelH97,DBLP:conf/oopsla/RenSTRC04,DBLP:conf/issta/GligoricEM15,10638618,DBLP:conf/icsm/ChesleyRR05,DBLP:journals/tse/FioravantiN01}. Existing techniques estimate impact using code structure and behavior~\cite{DBLP:conf/oopsla/RenSTRC04,nguyen2010recurring,DBLP:journals/stvr/LiSLZ13,10.1007/s10664-008-9088-2}, external artifacts such as change requests and execution data~\cite{DBLP:conf/iwpse/ShiriHR07,DBLP:conf/icsm/BriandLO03,10.1145/940071.940089,1553586}, and version-history patterns~\cite{zimmermann2006mining}. Other approaches combine multiple information sources for more accurate impact prediction~\cite{10.1109/SERA.2007.109,10.1109/WCRE.2010.21,yan2024enhancing}. Recent work further uses LLMs to capture developer intent and predict change impact~\cite{seed_to_scope}.

{CIA approaches focus on locating impacted locations, but do not suggest specific changes.
\tool can be viewed as a specialized change-impact-analysis workflow for coordinated renaming. It uses LLMs and static analysis, and leverages human feedback to suggest renames that align with the original change.}

\textbf{Identifier renaming} is one of the most frequent refactorings in modern codebases~\cite{jablonski2007cren, RENAS, renameEmpirical}. Many prior techniques focus on recommending improved names for individual identifiers~\cite{jablonski2007cren, abebe2013automated, dong2024context,zhang2023accurate,wang2025deep,liu2023refbert}, often integrated into IDEs for automated replacement.
% In contrast, our work targets \corename, which identifies semantically related identifiers that should be renamed together. 
Our approach complements these existing techniques by focusing on where renaming should propagate, not just what name to choose.

% {\bf Recommending Identifier Names.}
% Traditional methods use program analysis, rename histories, and lexical cues to suggest better names. For instance, CReN~\cite{jablonski2007cren} ensures consistent naming in clones, while CARER~\cite{dong2024context} enhances suggestions using static and dynamic context. Learning-based approaches treat naming as a language modeling task. Some rely on classifiers~\cite{zhang2023accurate} or deep models like RefBERT~\cite{liu2023refbert}, while others target naming inconsistencies~\cite{wang2025deep}. These tools complement to ours, which focuses on enforcing naming consistency across related identifiers when renaming.

{\bf Discovering \corename.} 
\corenameL focuses on identifying which identifiers should be renamed together. Empirical studies show that more than half of renames affect multiple elements~\cite{renameEmpirical}, motivating automated support.
RENAS~\cite{RENAS} uses the call-graph to extract relational knowledge, while RenameExpander~\cite{RenameExpander} uses structural proximity.
% decomposes edits and propagates changes using structural proximity. 
%These methods rely heavily on syntactic information.
% within the source code. 
{\tool} is designed to capture human intent,
% {(by declaring a refactoring scope)}
% and semantic alignment, 
enabling it to propagate renames even when syntactic cues are weak.
% {AlOmar et al.~\cite{alomar2024howToRefactor} show that developers often pair code fragments along with 
% natural-language descriptions of 
% refactoring intent when prompting LLMs -- this supports our design.}

{\bf Refactoring with LLMs and Agents.}
LLMs/agents have been used for suggesting method-level refactorings (e.g., extract/move method)~\cite{EM-Assist,MM-Assist,xu2025mantra}, code-smell removal~\cite{ismell}, and identifying modularity targets~\cite{localizeAgent}. 
% {Multi-agent systems like Mantra~\cite{xu2025mantra} and ISmell~\cite{ismell} perform refactorings such as smell removal or cohesion improvement. LocalizeAgent~\cite{localizeAgent} . 
% These systems 
% % pursue predefined goals, 
% pursue broad, high-level goals (e.g., localize a design issue), 
% whereas {\tool} addresses the subtler task of propagating abstract semantic changes inspired by the actions of the developer.}
These systems propose new edits, while {\tool} propagates renames already initiated by the developer. 
% These tools highlight the merit of enhancing traditional static analysis tools with LLMs~\cite{xie2025preMM,UTFix:OOPSLA12025,zhang2024pyDex,li2024enhancing} for covering a broader range of coding idioms used in real programs.

\textbf{Coding agents. }
% Unlike automatic program repair agents, e.g., AutoCodeRover~\cite{zhang2024autocoderover} or RepairAgent~\cite{bouzenia2024repairagent}, which aim to fix concrete bugs or optimize test outcomes, \tool handles abstract intent. It also enforces strict separation between reasoning and execution -- LLMs generate rename plans, but only IDE-native APIs perform edits, ensuring precision and safety.
SWE-agents possess generalist programming capabilities~\cite{anthropic2025claudecode,wang2025openhands,yang2024sweagent,zhang2024autocoderover,amazonqdev}.
Recent studies~\cite{gautam2025refactorbench,thillen_codetaste_2026,cordeiro_empirical_2026} have evaluated the capabilities of coding agents
% such as CluadeCode~\cite{anthropic2025claudecode} 
when performing complex refactoring tasks. RefactorBench \cite{gautam2025refactorbench} and CodeTaste~\cite{thillen_codetaste_2026} introduce benchmarks of multi-file refactoring tasks designed to assess stateful reasoning in agents. They find that SWE agents struggle with long edits, and required detailed prompts to achieve good results. 
%{Our experiments with ClaudeCode align with these findings (see Section \ref{sec:motivating}).}
Moreover, most coding agents operate in a request-and-response fashion that reduces the  developers' role to validate the agent’s output. However, our workflow is collaborative, and adapts continuously based on user feedback.
% \tool is a specific agent curated for \corename, which achieves SOTA performance.

% Kumar et al.~\cite{kumar2025sharptool} provide in-the-wild study of developer-agent collaboration, showing that successful outcomes hinge on iterative, incremental interaction between developer and SWE agents. These works underscore the challenges of applying agentic AI to real-world refactoring, further confirming our design of \tool as a developer-in-the-loop, multi-agent framework.}

%% file: Files/conclusion.tex
\section{Conclusion}
\label{sec:conclusion}

% \commentb{
% Software maintenance requires propagating concept changes across a codebase, demanding coordinated updates. Realizing this vision of coordinated editing in practice, however, is far from trivial.}

Our work shows that coordinated editing tasks, often viewed as routine maintenance, are fundamentally \emph{intent-propagation problems}. Successfully automating them requires explicitly representing the developer’s intended conceptual change, refining it through lightweight supervision, and propagating it consistently across a codebase.

We instantiate this paradigm through \tool, where the \dScope captures developer intent and is continuously refined during project-wide propagation. Rather than relying solely on stronger foundation models or unrestricted repository exploration, our workflow combines structured reasoning, trusted IDE-backed execution, and developer supervision to keep coordinated changes aligned with the intended conceptual change. A central finding is that developer supervision is essential not for discovering renames, but for refining the \dScope and preventing semantic drift, shifting developers from searching, remembering, and executing coordinated edits to supervising their propagation.

Beyond coordinated renaming, explicit intent propagation through a continuously refined \dScope provides a general blueprint for agent-assisted software engineering. 
%Many maintenance tasks require propagating conceptual changes across a project rather than performing isolated code transformations. 
We hope this work inspires the community to investigate broader classes of coordinated maintenance and software evolution tasks where developer intent can be explicitly represented, refined, and propagated through human–AI collaboration.

}

% \section*{Acknowledgements}

% We are grateful for the constructive feedback from the members of the AI Agents team at JetBrains Research. This research was partially funded through the NSF grants CNS-1941898, CNS-2213763, 2512857, 2512858, the Industry-University Cooperative Research Center on Pervasive Personalized Intelligence, and a gift grant from NEC.

% Add other friends who provided feedback:
% - Alejandra Garrido and her student Joaquin 
% - students in Danny's class who conducted writers' workshop: Blake, Ishika, Jarek, Vijay, Uditanshu
% - people from CUPLV at CU Boulder (via meeting that Abhiram had with them)
% Helena Klause from JetBrains AI Assistant